%% file: main.tex
\documentclass[12pt, a4paper, twoside]{article} 
\usepackage{a4wide}

\usepackage[small,labelfont=bf,hang]{caption}
\usepackage{graphicx} %%% for pdflatex use this line and comment out next
\usepackage{epsfig}
\usepackage{longtable}
\usepackage{colortbl}
\usepackage{lineno}
\usepackage{fancyhdr}
\usepackage{lipsum}%Dolor lipsum for simulation sec....
\usepackage{footnote}
\makesavenoteenv{tabular}

\usepackage{siunitx}

\usepackage{graphicx}
\usepackage{hyperref}
\usepackage{amsmath,amssymb}
\usepackage{booktabs}

\def\LALO          {LaAlO$_3$}

\def\MADMAX       {\mbox{{\sc Madmax}}}  

\def\etal       {{\it et al.}}

\newfont{\ord}{cmsy10 scaled 1200}

\begin{document}
%%%%%%%%%%%%%%%%%%%%%%%%%%%%%%%%%%%%%%%%%%%%%%%%%%%%%%%%%%%%%%%%%%%%%%%%%%%%%%%%
%
\begin{titlepage}
%
%%%%%%%%%%%%%%%%%%%%%%%%%%%%%%%%%%%%%%%%%%%%%%%%%%%%%%%%%%%%%%%%%%%%%%%%%%%%%%%%
%

\begin{center}
\includegraphics[scale=0.12]{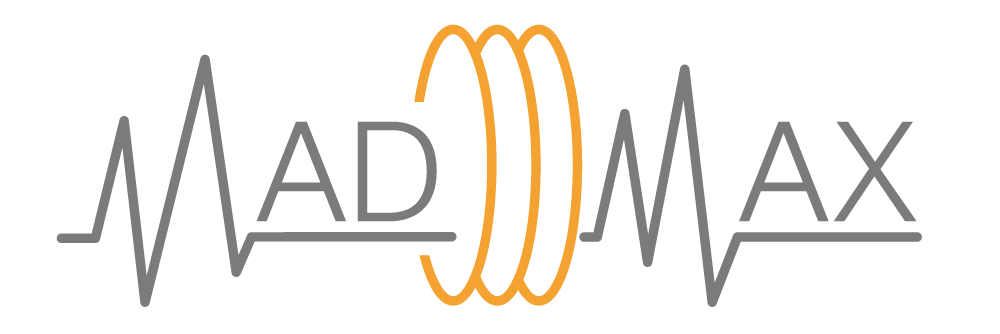}
%\vspace{0.5cm}

{\LARGE \bf Status report}
%\vspace{0.5cm}

{\bf  October 2019}
\end{center}
%
%\begin{figure}
%\hspace{-1.0cm}
%\vspace{0.5cm}
\centering
%\includegraphics[scale=0.3,angle=0]{FIGURES/2019-09-madmax-collaboration-logo-broad.jpg}
%\end{figure}

%\begin{figure}
%\vspace{-8truecm}

%\end{figure}
%\vspace{0.2cm}
%\centering The sensitivity plot of the \MADMAX\ white paper was chosen as  cover picture for EPJ C in the 2019 March edition.

%\clearpage
%\newpage
{\small%
\input{FIGURES/member_list.tex}
}

\vspace{1cm}

{\bf Abstract}

In this report we present the status of the MAgnetized Disk and Mirror Axion eXperiment (MADMAX),  the first dielectric haloscope for the direct search of dark matter axions in the mass range of 40 to 400 $\mu$eV.  MADMAX  will consist of several parallel dielectric disks, which are placed in a strong magnetic field and with adjustable separations. This setting is expected to allow for an observable emission of axion induced electromagnetic waves at a frequency between 10 and 100 GHz corresponding to the axion mass. The present document orignated from a status report to the DESY PRC in 2019.
\end{titlepage}

\newpage
\tableofcontents
\input{MADMAX_body.tex}

%\tableofcontents

%\bibliographystyle{authordate1}
%\bibliography{bibliography}

%\newpage

%%%%%%%%%%%%%%%%%%%%%%%%%%%%%%%%%%%%%%%%%%%%%%%%%%%%%%%%%%%%%%%%%%%%%%%%%%%%%%%%%%%%%%5
\end{document}

%% file: FIGURES/member_list.tex
\def\RWTH{\,{\bf $^{a}$}}
\def\MPR{\,{\bf $^{b}$}}
\def\NEEL{\,{\bf $^{c}$}}
\def\DESY{\,{\bf $^{d}$}}
\def\UHH{\,{\bf $^{e}$}}
\def\CPPM{\,{\bf $^{f}$}}
\def\MPP{\,{\bf $^{g}$}}
\def\CEA{\,{\bf $^{h}$}}
\def\TU{\,{\bf $^{i}$}}
\def\UZAR{\,{\bf $^{j}$}}
\begin{center}
%{\bf \Large The \MADMAX\ collaboration:}
%\vskip 5mm
\noindent
The \MADMAX\ collaboration: S.~Beurthey\CPPM{, }
N.~B{\"o}hmer\UHH{, }
P.~Brun\CEA{, }
A.~Caldwell\MPP{, }
L.~Chevalier\CEA{, }
C.~Diaconu\CPPM{, }
G.~Dvali\MPP{, }
P.~Freire\MPR{, }
E.~Garutti\UHH{, }
C.~Gooch\MPP{, }
A.~Hambarzumjan\MPP{, }
S.~Heyminck\MPR{, }
F.~Hubaut\CPPM{, }
J.~Jochum\TU{, }
P.~Karst\CPPM{, }
S.~Khan\TU{, }
D.~Kittlinger\MPP{, }
S.~Knirck\MPP{, }
M.~Kramer\MPR{, }
C.~Krieger\UHH{, }
T.~Lasserre\CEA{, }
C.~Lee\MPP{, }
X.~Li\MPP{, }
A.~Lindner\DESY{, }
B.~Majorovits\MPP{, }
M.~Matysek\UHH{, }
S.~Martens\UHH{, }
E.~{\"O}z\RWTH{, }
P.~Pataguppi\TU{, }
P.~Pralavorio\CPPM{, }
G.~Raffelt\MPP{, }
J.~Redondo\UZAR{, }
O.~Reimann\MPP{, }
A.~Ringwald\DESY{, }
N.~Roch\NEEL{, }
K.~Saikawa\MPP{, }
J.~Schaffran\DESY{, }
A.~Schmidt\RWTH{,}
J.~Sch{\"u}tte-Engel\UHH{, }
A.~Sedlak\MPP{, }
F.~Steffen\MPP{, }
L.~Shtembari\MPP{, }
C.~Strandhagen\TU{, }
D.~Strom\MPP{, }
G.~Wieching\MPR{, }
\end{center}
%\vskip5truemm
\begin{center}
{\rm \RWTH RWTH Aachen, Germany}\\[1mm]
{\rm \MPR Max-Planck-Institut f{\"ur} Radioastronomie, Bonn, Germany}\\[1mm]
{\rm \NEEL Institut NEEL, CNRF, Grenoble, France}\\[1mm]
{\rm \DESY DESY, Hamburg, Germany}\\[1mm]
{\rm \UHH Universit{\"a}t Hamburg, Hamburg, Germany}\\[1mm]
{\rm \CPPM CPPM, Aix Marseille Universite, CNRS/IN2P3, Marseille, France}\\[1mm]
{\rm \MPP      Max-Planck-Institut f{\"ur} Physik, M{\"u}nchen, Germany}\\[1mm]
{\rm \CEA CEA-Irfu Saclay, France}\\[1mm]
{\rm \TU   Eberhard-Karls-Universit{\"a}t T{\"u}bingen, T{\"u}bingen, Germany}\\[1mm]
{\rm \UZAR Universidad Zaragoza, Zaragoza, Spain}\\[1mm]
%{\rm}\\[1mm]
%{\rm}\\[1mm]

\vspace*{5mm}
%\begin{center}
%\parbox{65mm}{\hfill\underline{Spokesperson}:~~ }\parbox{100mm}{
% B. Majorovits~~({\it bela$@$mpp.mpg.de})}\\[1mm] 
%\parbox{65mm}{\hfill\underline{Chair of Collaboration Board}:~~ }\parbox{100mm}{
% E. Garutti~~({\it Erika.Garutti$@$physik.uni-hamburg.de})}\\[1mm]   
%\parbox{65mm}{\hfill\underline{Physics Coordinator}:~~ }\parbox{100mm}{
% A. Schmidt~~ ({\it Alexander.Schmidt$@$physik.rwth-aachen.de})}\\[4mm]

%\parbox[b]{80mm}{
%\begin{tabular}{rl}
%chair of speakers bureau:& J. Jochum/K. Zuber\\
%chair of editorial board:& P. Grabmayr
%\end{tabular}
%}
 
\vspace*{5mm}
\parbox{45mm}{\hfill URL:~~ }\parbox{100mm}{\tt http://madmax.mpp.mpg.de/}          
 
%\end{center}

\end{center}

%% file: MADMAX_body.tex
% -*- coding:utf-8 -*-
% vi:encoding=utf-8:
% !TEX encoding = UTF-8 Unicode
% This file should be utf8 encoded so that these characters render as
% umlauts: ÄÖÜäöüß
%

%All sections should include comments, if applicable, on
%milestones and decision points as well as lessons we would like to learn from prototyping!

\input{ExecutiveSummary.tex}

\newpage
\section{Physics motivation and foundation}
\label{sec:physics_motivation}

\subsection{The case for axions}
Searches for a neutron electric dipole moment $d_\mathrm{n}$ find ${|d_\mathrm{n}|<2.9\times 10^{-26}e\,\mathrm{cm}}$~\cite{Baker:2006ts,Abel:2020gbr,Afach:2015sja} with electron charge $e$. Together with limits on other nuclear electric dipole moments, this implies an absence of charge-parity (CP) violation in the strong interactions at least to a degree that is highly puzzling, in particular, in light of the existing sizable CP violation in the electroweak sector of the standard model (SM) of particle physics. The absence  of measurable CP violation is one of the most puzzling features within the SM, also called the strong CP problem.

An elegant solution to this puzzle was proposed by Peccei and Quinn in the year 1977~\cite{Peccei:1977hh,Peccei:1977ur}: By extending the SM with a new global chiral U(1) symmetry---the Peccei--Quinn (PQ) symmetry U(1)$_\mathrm{PQ}$---that is broken spontaneously at the PQ scale $f_a$, a mechanism is introduced, which suppressed dynamically CP violation in the strong interactions. The axion, $a$ emerges 
as the associated pseudo-Nambu-Goldstone boson~\cite{Weinberg:1977ma,Wilczek:1977pj} with a mass today given by
$m_a\simeq 5.7
\,\mu\mathrm{eV}(10^{12}\,\mathrm{GeV}/f_a)$~\cite{diCortona:2015ldu}.

While the original PQ proposal assumed $f_a$ to be at the weak scale, axion searches, astrophysical observations and cosmological arguments point to a very high scale of $f_a\gtrsim 3\times 10^{8}\,\mathrm{GeV}$ implying a very small axion mass of $m_a\lesssim 0.02~\mathrm{eV}$~\cite{raffelt,Chang:2018rso,Giannotti:2017hny}.

Remarkably, the axion is also an excellent cold dark matter (CDM) candidate~\cite{Preskill:1982cy,Abbott:1982af,Dine:1982ah}. The $m_a$ range in which the correct CDM abundance is expected to be provided by the axion depends on the order of two critical events in the past: cosmic inflation and the PQ symmetry breaking.

In the first scenario, PQ symmetry breaking occurred before inflation without any subsequent PQ symmetry restoration. Thereby, the initial value of the axion field in our local universe is unique ---up to quantum fluctuations--- and is fundamentally unpredictable from first principles. The vacuum realignment mechanism~\cite{Preskill:1982cy,Abbott:1982af,Dine:1982ah} can then provide the complete amount of CDM in the form of cold axions for any value of $f_a\gtrsim 10^{10}\,\mathrm{GeV}$ corresponding to a large possible region of the axion CDM mass of $m_a \lesssim 0.5~\mathrm{meV}$. The exact value of the axion CDM mass, however, cannot be theoretically predicted because of the random initial value of the axion field in the observable universe.

In the second scenario, inflation occurred before PQ symmetry breaking. This implies a patchy structure of the axion field in the observable universe with its initial value being essentially random in each patch of the universe causally disconnected during PQ symmetry breaking. The relic axion CDM density from the vacuum realignment mechanism is then given by the statistical average and depends on $f_a$ only~\cite{Preskill:1982cy,Abbott:1982af,Dine:1982ah}. However, the patchy structure is associated with cosmic strings and domain walls which complicates the calculation of the axion CDM density such that the calculation of the exact value of the axion CDM mass --- i.e.\ the $m_a$ value for which axions provide the correct CDM density --- in this scenario is still the subject of frontier research in astroparticle physics~\cite{1,2,Ringwald:2015dsf,Fleury:2015aca,Fleury:2016xrz,Moore:2016itg,Klaer:2017ond,Gorghetto:2018myk}. Currently, the mass range $26\,\mu{\rm eV}\lesssim m_a \lesssim 1\,{\rm meV}$
is considered as the best motivated one in the post-inflationary PQ-symmetry-breaking scenario with $m_a\sim 100~\mu$eV as a typical representative value for the CDM axion mass.

\subsection{The dielectric haloscope}
%As a dielectric haloscope, \MADMAX{} will search for a microwave signal sourced by the axion CDM field in a strong magnetic dipole field $\mathbf{B}_\mathrm{e}$. 
The underlying fundamental physics of a dielectric haloscope is described by the modified Maxwell equations (cf.~\cite{madmax_foundations} and references therein) obtained when accounting for the Lagrangian density describing axion-photon interactions, 
\begin{equation}
\mathcal{L}_{\mathrm{int}}
=
-\frac{\alpha}{2\pi}C_{a\gamma}\mathbf{E}\cdot\mathbf{B}\,\theta,
\label{eq:Lint}
\end{equation} 
where $\mathbf{E}$ and $\mathbf{B}$ are electric and magnetic fields
%, $\theta=a/f_a$ 
and $\alpha=e^2/(4\pi)$ the fine-structure constant in the Lorentz-Heaviside convention and natural units with $\hbar=c=1$. 
%
%As can be seen in~\eqref{eq:Lint}, 
The axion-photon interaction strength is governed by the axion-model-specific parameter $C_{a\gamma}=\mathcal{E}/\mathcal{N}-1.92\equiv (2\pi/\alpha)\,f_a\,g_{a\gamma}$ 
which is usually of $\mathcal{O}(1)$ and given by the electro-magnetic (EM) and color anomalies of the PQ symmetry, $\mathcal{E}$ and $\mathcal{N}$, respectively~\cite{diCortona:2015ldu}.
%where the number in brackets refers to the uncertainty in the last digit~\cite{diCortona:2015ldu}.
%
Considering a medium with permeability $\mu=1$ and a dielectric constant $\epsilon$ inside of an external static and homogeneous $B$-field $\mathbf{B}_\mathrm{e}$, the corresponding modified Maxwell equations imply the existence of a tiny axion-induced electric field
\begin{equation}
\mathbf{E}_a(t)
=
-\frac{\alpha}{2\pi\epsilon}C_{a\gamma}\mathbf{B}_\mathrm{e}\theta(t).
\label{eq:Eaxioninduced}
\end{equation} 
At an interface between media with different $\epsilon$, $\mathbf{E}_a(t)$ is discontinuous, whereas the (modified) Maxwell equations require continuous total $\mathbf{E}$ (and $\mathbf{H}$) field components parallel to the boundary. The continuity is ensured by the emissions of electromagnetic radiation at the boundary in the direction perpendicular to the boundary.

This effect has been discussed first as the dish antenna concept~\cite{dish} and is used by the BRASS project pursued at the University of Hamburg~\cite{BRASS}. The corresponding power output of a single magnetized metallic mirror of area $A$ is~\cite{dish}
\begin{eqnarray}
P_{\gamma,0}
= 2.2\times 10^{-27}\,\mathrm{W}
\left(\frac{A}{1~\mathrm{m}^2}\right)\!
\left(\frac{B_{\rm e}}{10~{\rm T}}\right)^{\!\!2} \!
\left(\frac{\rho_a}{0.3~\mathrm{GeV}/\mathrm{cm}^3}\right)
C_{a\gamma}^2 \, ,
\label{eq:PMirror}
\end{eqnarray}
where $\rho_a$ is the galactic local axion CDM density.
While the emission from the mirror surface would be frequency independent, it seems technologically difficult to magnetize a sufficiently large surface area $A$ with a high enough $\mathbf{B}_\mathrm{e}$-field perpendicular to the surface to obtain a detectable power output induced by the axion CDM field.

Dielectric haloscopes are using  movable dielectric discs in front of (or between) metallic mirror(s) exploiting constructive interference and resonant enhancements of the radiation emitted at the many interfaces. A related concept has been discussed earlier~\cite{dielectric_lbl} and a first proof of principle setup, ORPHEUS, has obtained first results \cite{orpheus}.

\begin{figure}
    \centering
    \includegraphics{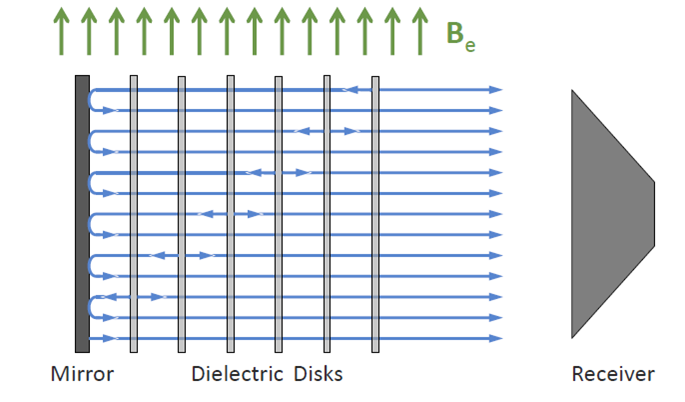}
    \caption{A dielectric haloscope consisting of a mirror and several dielectric discs placed in an external magnetic field ${\bf B}_{\rm e}$ and a receiver in the field-free region. A focusing mirror (not shown) could be used to concentrate the emitted power into the receiver. Internal reflections are not shown. From~\cite{madmax_prl}.}
    \label{fig:madmax_principle}
\end{figure}

The \MADMAX{} experiment is based on a similar idea~\cite{dielectric,madmax_prl,madmax_foundations}: 
%
%The coherent emission at all the surfaces can lead to resonant enhancements of the signal for frequencies that are adjustable by the spacings between the individual discs. 
%
Fig.~\ref{fig:madmax_principle}
illustrates a dielectric haloscope consisting of one mirror and six dielectric discs placed in an external magnetic field $\mathbf{B}_\mathrm{e}$ (green) parallel to the disc surfaces, and an antenna coupled to a receiver in the field-free region. The photon emissions (ignoring internal reflections) is sketched by blue lines.
%
%The basic principle of the dielectric axion to photon boost is visualized in Fig.\,\ref{fig:madmax_principle}.
%

\subsection{Sensitivity reach of the dielectric haloscope}
\begin{figure}[h!]
    \centering
    \includegraphics[width=0.9\linewidth]{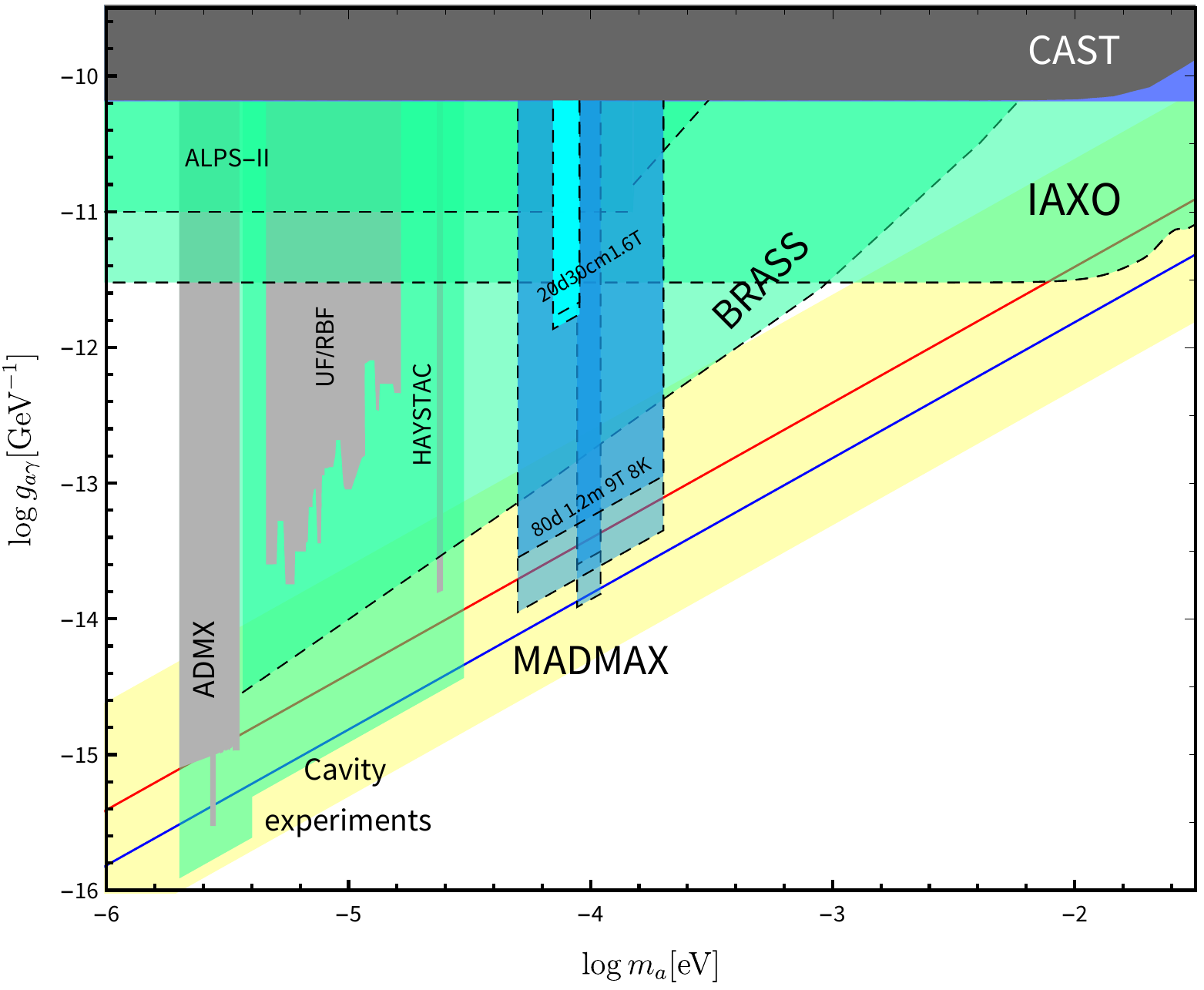}
    \caption{Sensitivity  of \MADMAX{} shown in the parameter region spanned by the axion/ALP mass $m_a$ and the axion/ALP-photon coupling $g_{a\gamma}$. The diagonal red and blue lines indicate the location of two prominent QCD axion models (KSVZ and DFZS, respectively). The yellow area displays the range of plausible models for the ratio $\mathcal{E}/\mathcal{N}$ \cite{diLuzio}. Astrophysical limits (blue) and limits from searches (gray) with cavity haloscopes (ADMX, HAYSTAC, UF/RBF \cite{uf,rbf}) and the axion helioscope CAST are shown. The expected sensitivity of future cavity experiments (e.g.\ ADMX, HAYSTAC, CULTASK), the axion helioscope IAXO, the light-shining-through-the-wall experiment ALPS-II and the magnetized dish antenna haloscope BRASS are also indicated. The sensitivity range for \MADMAX{} is shown for two cases (blue regions): a wide- and a narrow-mass-range scan.
    %with 50~MHz and 20\,MHz boost-factor bandwidth for 3 and 5 years scanning time, respectively. 
    For both cases the sensitivity is shown for a "conservative" and a more "success oriented" set of parameters (see text).
     Also the projected sensitivity for the prototype with 20 discs in a 1.6\,T field at 8K within 90 days measurement time is shown in cyan (see Sec.\,\ref{sec:prototype}). \MADMAX{} sensitivities correspond to 5\,$\sigma$ signal above noise. The other limits and sensitivities may have other definitions.}
    \label{fig:sensitivity_plot}
\end{figure}

With the proposed dielectric haloscope \MADMAX{} it is planned to detect CDM axions if they are in the mass range 
\begin{equation}
40\,\mu{\rm eV}\lesssim m_a \lesssim 400\,\mu{\rm eV} . %\mu{\rm meV}
\label{eq:massrangeMADMAX}
\end{equation} 
Sensitivity to both the KSVZ \cite{KSVZ} and DSFZ \cite{DSFZ} QCD axion models in this range will allow for tests of both axion CDM scenarios: 
The pre-inflationary PQ-symmetry-breaking scenario for relatively large initial misalignment angles of $\theta_i\gtrsim2.5$ and a significant fraction of the mass region motivated by the post-inflationary PQ-symmetry-breaking scenario.
Thereby, \MADMAX{} will be highly complementary to cavity halosopes searching for a resonantly-enhanced signal induced by the CDM axion in a cavity placed in a strong magnetic field (Sikivie's haloscopes~\cite{sikivie}) such as ADMX~\cite{admx}, ADMX HF~\cite{admx-hf}, HAYSTAC~\cite{Zhong:2018rsr} or CULTASK~\cite{Chung:2018wms}. Cavity haloscopes are the most sensitive axion CDM searches today and seem to be optimal for $m_a\lesssim 40\,\mu\rm eV$, i.e., below the \MADMAX{} search range~\eqref{eq:massrangeMADMAX}. 
For larger $m_a$ values, smaller cavities are required to allow for the CDM axion-induced resonance which is in turn associated with substantial challenges to obtaining a signal sensitivity at an observable level.
In contrast, existing studies suggest that \MADMAX{} has the potential to cover an important range in the particularly well-motivated axion CDM mass region for the first time with a sensitivity expected to reach the most prominent QCD axion models (see Fig.~\ref{fig:sensitivity_plot}).

\MADMAX{} will aim to detect the CDM axions bound to our galaxy, which are assumed to provide the full local galactic CDM density of $\sim0.3~\mathrm{GeV}/\mathrm{cm}^3$~\cite{{Tanabashi:2018oca}} and to have a velocity dispersion on Earth described by the galactic virial velocity of~$v_a\sim10^{-3}\,c$. The associated macroscopic de Broglie wavelength of axions $\lambda_{\mathrm{dB}}=2\pi/(m_{a}v_{a})=12.4~\mathrm{m}\,(100~\mu\mathrm{eV}/m_a)(10^{-3}\,c/v_a)$ will exceed the size of the \MADMAX{} booster system significantly in particular towards smaller $m_a$ values.
%
%at least an order of magnitude. 
%
Thus, we expect \MADMAX{} to probe the axion CDM field $a=\theta f_a$  as an (approximately) homogeneous and monochromatic classical oscillating field $\theta(t)\propto \theta_0\cos(m_a t)$ with $\theta_0\sim 4\times 10^{-19}$ fixed by the local CDM density, $\rho_a= f_a^2 m_a^2 \theta_0^2/2\sim 0.3~\mathrm{GeV}/\mathrm{cm}^3$, and a frequency of $\nu=m_a/(2\pi)$ in the microwave range of $10~\mathrm{GHz}\lesssim \nu \lesssim 100~\mathrm{GHz}$.

%Without the dielectric discs, 
%
By optimizing the spacings between the individual discs (each of transverse surface area $A$), it can be shown that a significant frequency-dependent boost of the power output can be achieved: %for certain frequency ranges:
\begin{eqnarray}
P_{\gamma}(\nu)
= \beta^2(\nu)\,P_{\gamma,0}
= 1.1\times 10^{-22}\,\si{\watt} \left(\frac{\beta^2(\nu)}{5\times 10^4}\right)\! 
\left(\frac{A}{1\,\mathrm{m}^2}\right)\!
\left(\frac{B_{\rm e}}{10\,{\rm T}}\right)^{\!\!2}\! 
\left(\frac{\rho_a}{0.3\,\mathrm{GeV}/\mathrm{cm}^3}\right) 
C_{a\gamma}^2 \, .
\label{eq:PDielectricHaloscope}
\end{eqnarray}
Here the enhancement as a function of frequency with respect to a single magnetized mirror is quantified in terms of the power boost factor, $\beta^2(\nu)$, generally referred to as boost factor in the following. It depends on the widths, separations, and on the dielectric constant $\epsilon$ of the dielectric discs.

\begin{figure}[t]
\begin{center}
\includegraphics[width=7.5cm,height=6.5cm]{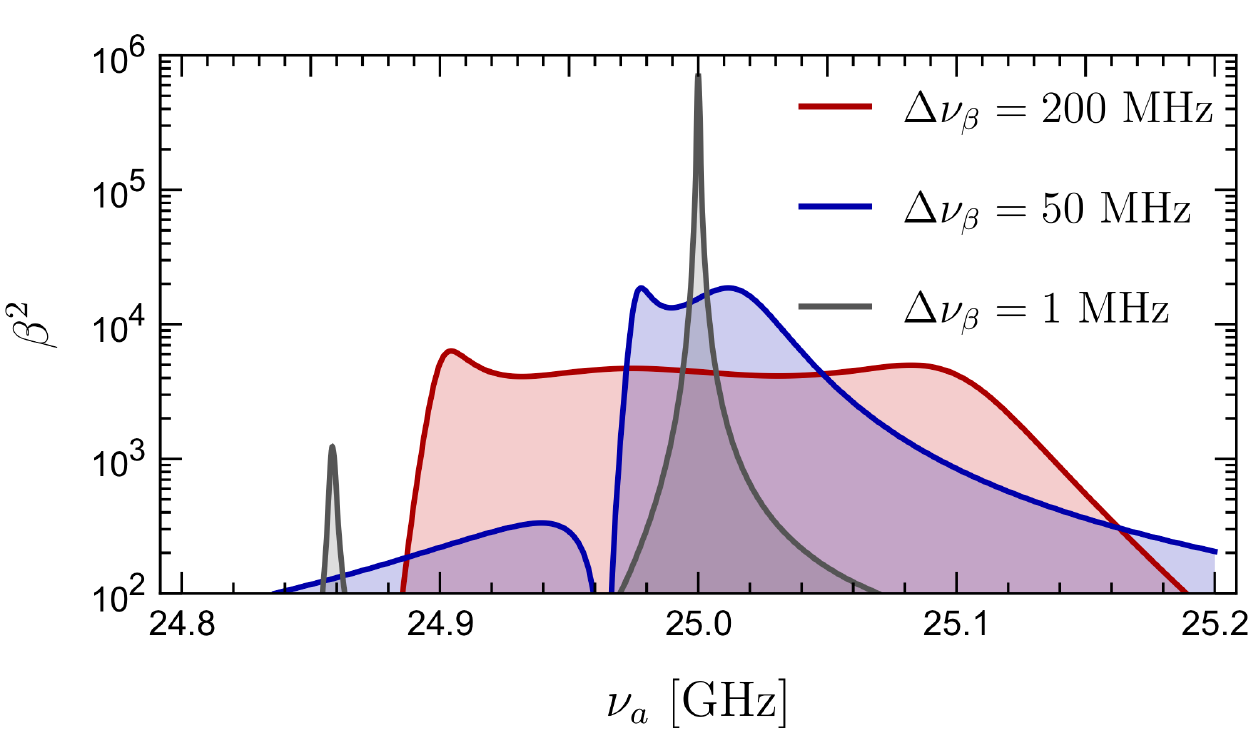}
\hfill
\includegraphics[width=7.5cm,height=6.8cm]{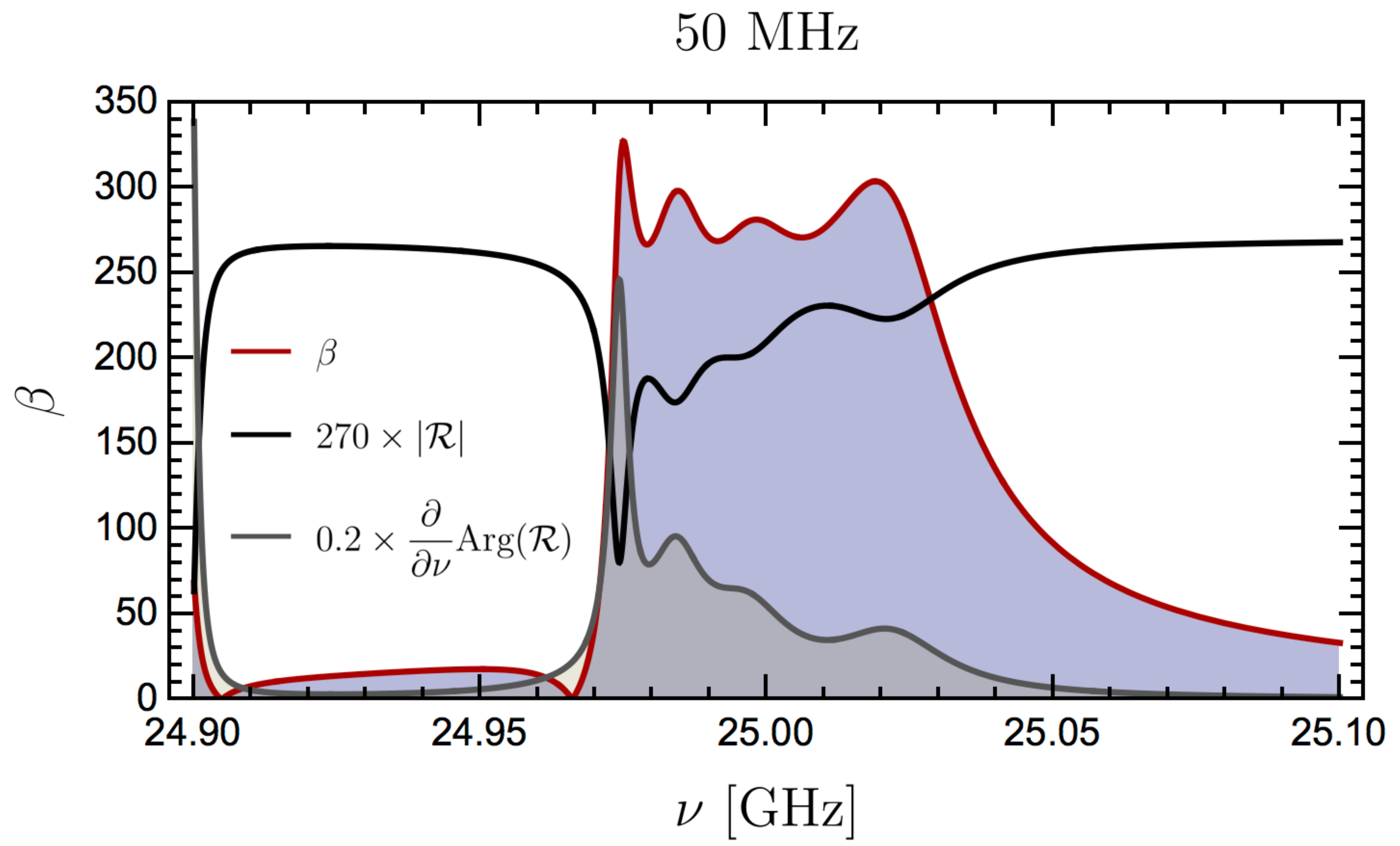}
\caption{
\textbf{(left)}~Boost factor $\beta^2(\nu)$ for 20~discs ($d=1~\mathrm{mm}$, $\epsilon=25$) optimized for $\Delta\nu=200$, 50, and 1~MHz (red, blue, and gray) centered on 25~GHz. Adapted from~\cite{madmax_prl}. 
\textbf{(right)}~Amplitude boost factor $\beta(\nu)$ (red) optimized for $\Delta\nu=50$~MHz centered on 25~GHz and scaled values of reflectivity $|{\cal R}(\nu)| $ (black) and group delay $\frac{\partial}{\partial \nu}\text {Arg}[\mathcal R(\nu)]$ (gray) for 80~discs ($d=1~\mathrm{mm}$, $\epsilon=25$). The reflectivity is obtained for lossy dielectric discs with a loss tangent of $\tan\delta\sim\,$few$\times 10^{-4}$. 
%and more specifically a refractive index $n$ with ${\rm Im}(n)=10^{-3}$. 
From~\cite{Millar:2017eoc}.
}
\label{fig:powerboostNboost}
\end{center}
\end{figure}

Fig.~\ref{fig:powerboostNboost}~(left) shows $\beta^2(\nu)$ optimized for maximum values within the frequency ranges $\Delta\nu=1$, 50 and 200~MHz, each equally centered on 25~GHz for a dielectric haloscope consisting of 20 magnetized discs (\hbox{$d=1~{\rm mm}$}, \hbox{$\epsilon=25$}) and a metallic mirror. 
The curves are obtained for an idealized 1-dimensional setting with a mirror and discs of infinite transverse extent and perfect parallel alignment. 
The figure illustrates that already 20 discs allow for significant $\beta^2$ values and that an area law holds according to which $\int d\nu \beta^2$ is preserved, such that it is possible to trade width for power and vice versa~\cite{madmax_foundations}. 
%
%Moreover, it turns out that $\int d_{\nu_a}\beta^2$ is proportional to the sum over interfaces, which holds exactly when integrating over $0\leq\nu_{a}\leq\infty$. 
%
In addition, the area under the $\beta^2$ curve is approximately proportional to the number of discs $N$. Results for realistic 3-dimensional settings of finite extent will be presented in section~\ref{sec:sim}. 
%
%Based on the latter and on~\eqref{eq:PoverADielectricHaloscope}, the design goals of MADMAX---aiming at a sensitivity allowing us to probe QCD axion models---are a dielectric haloscope consisting of a mirror and 80~dielectric discs of about $1~\mathrm{m}^2$ transverse area each placed inside a magnetic dipole field of about $10~\mathrm{T}$.

%It is important to stress that the booster power output of the system appears only in the presence of the CDM axion (corresponding ALP or hidden photon), i.e., at $\nu_a\simeq m_a$. Thus, w

When searching for the axion, the central $\nu$ value around which the $\Delta\nu$ region with significant $\beta^2$ is expected has to be seamlessly shifted to cover the mass range in~\eqref{eq:massrangeMADMAX}. 
To verify that the disc spacings are set properly to allow for the significant $\beta^2$ value in the considered frequency range, the reflectivity $|{\cal R}(\nu_a)|$ and the group delay $\frac{\partial}{\partial \nu_a}\text {Arg}(\mathcal R)$ of the system will be crucial quantities. As shown in Fig.~\ref{fig:powerboostNboost}~(right) both show features related to the amplitude boost factor $\beta$ at $\nu=m_a/(2\pi)$, but are measurable at other frequencies as well. Indeed, these quantities are being studied in the laboratory as will be discussed in section~\ref{sec:meas-setup}.
The curves in Fig.~\ref{fig:powerboostNboost}~(right) are obtained for the mentioned 1-dimensional idealized setting, but now for 80~discs each of thickness $d=1~\mathrm{mm}$ and $\epsilon=25$, where a loss tangent of $\tan\delta\sim\,$few$\times 10^{-4}$ is assumed in the calculation of the reflectivity.

While the final \MADMAX{} setup should ultimately be sensitive to QCD axion models, there is a significant unexplored parameter range of Axion Like Particles (ALPs) and of hidden photons that should be accessible with the down-scaled \MADMAX{} prototype (discussed in section~\ref{sec:prototype}). 
%as a down-scaled dielectric haloscope. 
In the case of hidden photons, an unexplored parameter range would be accessible even without the external magnetic field $\mathbf{B}_\mathrm{e}$. While neither ALPs nor hidden photons are associated with a solution of the strong CP problem, both could be present in SM extensions 
%of the standard model of particle physics 
and could explain the CDM in our universe. %As such they could be accessible to the \MADMAX{} prototype. 

Fig.~\ref{fig:sensitivity_plot} shows the parameter region spanned by the axion-ALP mass $m_a$ and the axion-ALP-photon coupling $g_{a\gamma}$ with astrophysical limits (blue), limits from searches (gray), projected sensitivities of planned projects (green) and the location of typical QCD axion models (red and blue lines).
Presently, such models \cite{KSVZ, DSFZ} in the axion CDM mass range are tested only by the ongoing ADMX cavity haloscope searches, and by the solar axion search with the helioscope CAST towards relatively large $m_a$ values~\cite{Anastassopoulos:2017ftl}.

The discovery potential of \MADMAX{} shown in Fig.\,\ref{fig:sensitivity_plot} is expected to touch the QCD axion model line as are planned cavity haloscopes such as CULTASK~\cite{Chung:2018wms} and the axion helioscope IAXO~\cite{Armengaud:2019uso}. Also, the discovery reach of the light-shining-through-the-wall experiment ALPS-II~\cite{ALPSII} and the magnetized dish antenna experiment BRASS~\cite{BRASS} are shown in Fig.~\ref{fig:sensitivity_plot}.
The  sensitivity of the \MADMAX{} project and the period needed to scan the full mass range depends on the assumptions on achievable boost factor (loss effects), achievable noise temperature of the receiver ($T_{\rm sys}$), coupling of the antenna-receiver system to the booster and the readjustment time ($t_R$) for the disc positions between measurements at different mass-ranges. The depicted sensitivity reach of \MADMAX{} is defined as a signal with 5\,$\sigma$ above system noise temperature. 
It is assumed that 
%the coupling between the axion-induced EM field and the receiver is 
50\,\% of  the theoretically obtainable maximum power is detected (30\% loss from 3D effects, and another 30\% from antenna coupling). 
The reach of \MADMAX{} is shown for a wide- and a narrow-mass-range scan,  each of them under two sets of assumptions: ``conservative'' (upper dashed line) and ``success oriented'' (lower dashed line). The conservative set assumes boost-factor bandwidth $\Delta\nu_\beta = 50$ MHz,  $t_R=1$ day, $T_{\rm sys}=8$ K and a live time of 3 years. The success oriented assumes $\Delta\nu_\beta = 20$ MHz, $t_R=$ 1h, $T_{\rm sys}=4$ K and 5 years scanning time. 

%These will be discussed in some details below. 

%EG %\section{\MADMAX{} concept, collaboration and working group structure}

\section{\MADMAX{}: concept, collaboration and strategy}
\label{sec:concept}
\subsection{The \MADMAX{} concept}

The \MADMAX{} project is based on the idea of the dielectric haloscope:
%as described in section~\ref{sec:physics_motivation}. 
the axion induced power output is boosted by the system to $\approx$\,10$^{-22}$\,W, cf.~eq.~\eqref{eq:PDielectricHaloscope}. This can be realistically  measured with high enough significance within a $\sim$ week of integration time with an overall noise temperature of 8\,K. To obtain this power, a power boost $\beta^2\gtrsim\,2\times 10^4$ and a dipole magnet with a  figure of merit 
\begin{equation}
    \mathrm{FoM}=\frac{1}{L}\int_A \int_0^L B(x,y,z)^2 dz dx dy \approx 100~\mathrm{T}^2 \mathrm{m}^2,
\end{equation}
are aimed for.
Here $B(x,y,z)$ gives the $y$-component of the $B$-field as a function of the position,
%with inhomogeneities of the B-field $\lesssim$\,20\%, 
$L$ is the maximum length of the booster\footnote{The $z$-axis points along the axis of the cylindrical booster. The $y$-axis is vertical.} and, $A$ is the area of each disc contributing to the axion to photon conversion.

\begin{figure}
    \centering
    \includegraphics[width=1.0\linewidth]{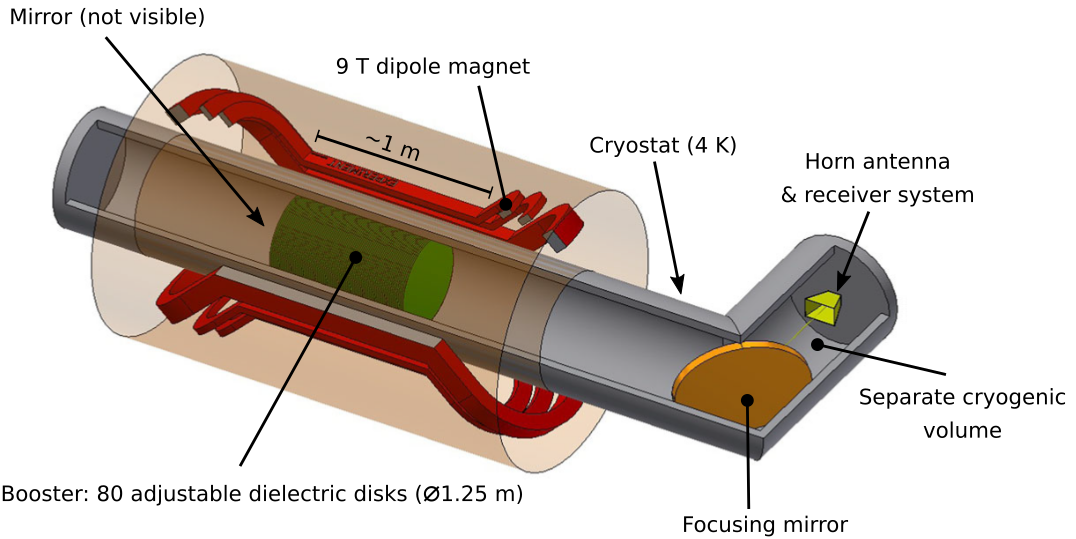}
    \caption{Conceptual sketch of the \MADMAX{} experiment (taken from \cite{madmax_whitepaper}).}
    \label{fig:madmax_concept}
\end{figure}

The planned \MADMAX{} experiment would consist of the following main components: 
\begin{itemize}
    \item {\bf Magnet}: A $\approx$\,9\,T dipole magnet with a 1.35\,m warm bore out of which the inner 1.2\,m should be available for the disc areas to get  FoM\,=100\,T$^2$m$^2$;
    \item {\bf Booster}: The booster including $\sim$\,80  dielectric discs in front of a mirror with $\approx$\,1.2\,m$^2$ area, adjustable in their relative distance between 1.5 and 15\,mm with a precision of better than $\approx$ 10~$\mu$m;
    \item {\bf Receiver}: An exchangeable receiver to enable the detection of signal photons in the frequency range of 10-100\,GHz coupled to the optical system that is guiding the power output of the booster into the receiver.
\end{itemize}
 The booster and optical system will be situated in a liquid helium cooled cryostat, in order to reduce the system noise temperature. The receiver will be housed in a separate cryostat. A conceptual sketch of such a system (without the receiver cryostat) is shown in Fig.~\ref{fig:madmax_concept}.

\subsection{The \MADMAX{} collaboration}
The \MADMAX{} collaboration was founded at DESY-Hamburg in October 2017. Currently, it consists of eight institutes as full members and two as associate members. 
The member institutes are listed in Tab.~\ref{tab:members}.

\begin{table}
\begin{tabular}{l|l}
{\bf Full members} & Acronym \\
\hline
RWTH Aachen, Germany  & RWTH \\
CEA-Irfu Saclay, France & CEA\\
DESY Hamburg, Germany & DESY \\
Universit{\"a}t Hamburg, Germany  & UHH\\
Max-Planck-Institut f{\"ur} Physik, M{\"u}nchen, Germany & MPP \\
Max-Planck-Institut f{\"ur} Radioastronomie Bonn, Germany & MPIfR \\
Eberhard-Karls-Universit{\"a}t T{\"u}bingen, Germany & Uni. T{\"u}bingen\\
Universidad Zaragoza, Spain & Uni. Zaragoza\\
\hline
\hline
{\bf Associate members } & Acronym  \\
\hline
Centre de Physique des Particules de Marseille, France &  CPPM\\
Institut NEEL, CNRS, Grenoble, France & NEEL \\
\end{tabular}
\caption{List of \MADMAX{} member institutions}
\label{tab:members}
\end{table}

%Presently the \MADMAX{} collaboration is organized into working groups, covering the main tasks of understanding the detailed RF response of the system,  development of the hardware components, prototype integration, and of providing the magnet and the infrastructure at DESY/CERN. 
%The coverage of work packages is listed in Tab.\,\ref{tab:WP}. 

%\begin{table}[h]
%    \centering
%    \begin{tabular}{l|l}
%{\bf Work package } & {\bf Institutes involved} \\
%\hline
% Theory, simulations & Uni. Zaragoza, RWTH\\
% Magnet & MPP, DESY\\
% Infrastructure at DESY \& cryogenics & DESY\\
%\hline
%%{\bf Prototype tasks} & \\
%\hline
% Proof of principle booster & MPP\\
% Cryo-engineering of booster & UHH, MPP, CPPM\\
% Optical system components & MPIfR \\
% Tiling \& characterization of discs & UHH, RWTH, CPPM \\
% Interferometer procurement and implementation & Uni. T\"ubingen\\
%Cryostat specification, procurement, and commissioning & UHH, DESY \\
%Prototype booster  building and commissioning & DESY, UHH, MPP, RWTH\\ 
% Receiver & MPP, NEEL \\
% Infrastructure at CERN & CPPM \\
% Infrastructure for prototype & UHH \\
% Data analysis \& computing & MPP \\
%    \end{tabular}
%    \caption{Work packages definition and task sharing between institutes.  }
%    \label{tab:WP}
%\end{table}{}

\subsection{R\&D strategy}
The project includes several technological challenges.
They will all (except for the magnet) be addressed by a scaled-down version of the final experiment -- the prototype.
The output of the tests with the prototype will be essential for scaling up for the final design of the  \MADMAX{} experiment.
The key challenges are detailed below together with the strategy to tackle them.

\subsubsection{Magnet} 
\label{sec:magnet_challenges}
A dipole magnet with a FoM\,$\approx$\,100\,T$^2$m$^2$ is necessary. 
This implies that the stored energy would be $\sim$\,480 MJ, that the peak field inside the superconductor would be close to the critical field of NbTi, and that the coils and supporting structures would need to withstand strong forces.
No dipole magnet with such challenging specifications has ever been built. It was
not clear at the beginning of the project whether this could be built within reasonable budget and time scale.
Feasibility, time scale, and costs are strongly interlinked. To tackle this in the most efficient way, an innovation partnership with two competing magnet suppliers started (CEA-Irfu and Bilfinger Noell) in 2017. %It is the first one within the Max Planck society (MPG) and has the full support of the general administration of the  MPG. 

In the first phase the two  partners performed independent feasibility studies. The goal was to show that it would be possible to build a magnet with the specified FoM\,=\,100\,T\,$^2$m$^2$ using NbTi superconductor. Both innovation partners came up with a similar design and consistent conclusions. The results were regularly evaluated and the plans to move forwards were approved by an external expert committee \footnote{R. Gupta - BNL, D. Sellmann -DESY, H. ten Kate - CERN, S. Stoynev - Fermilab, F. Toral - CIEMAT}. 

%In order to evaluate the feasibility of building such a magnet within a budget and time scale believed to be realistic,  

%The project is structured into project phases:
%%phases similar to the overall project: 
%feasibility, demonstrator and final magnet. 

%Within the frame of the innovation partnership, all legal tendering issues until final delivery of the magnet have already been clarified. For instance, one key advantage is that no tendering process will have to be performed by the institutes. This construct allows the collaboration to plan the money flow flexibly.

Since finishing the initial competitive feasibility study, the innovation partners have joined forces and are cooperating towards the realization of the magnet. This allows for the most efficient exploitation of the complementary expertise of the academic and the industrial partner.  
%The close cooperation between the partner allows the \MADMAX{} collaboration to take over risks that would otherwise have to be payed for. Like this the the overall price for the magnet can be reduced significantly. 

The partnership structure is managed together with representatives of the collaboration. %The coordinator of the magnet project is appointed by the Collaboration Board. 

\subsubsection{Determination of the boost factor}
The strength of an expected axion signal will be governed by the boost factor of the setup in the  frequency range under consideration. It is therefore very important to properly understand the boost factor and the coupling of the signal into the detector system. A direct measurement of the boost factor can not be  made. It is therefore essential to develop an in-depth understanding of the response of the system. This will be done by developing complementary methods to measure and simulate the RF response. Possibilities of in situ calibrations of the system response using  transmission, reflectivity and group delay  are described below (see Sec.\,\ref{sec:meas-setup}). Some promising ideas to measure the boosting behavior, like enhanced thermal noise of the booster discs \cite{thermal_noise} or emission of radiation from surface currents on the mirror, are presently being investigated.
These measurements allow indirect calibration of the boost factor. Comparison with simulations will help to verify that the RF behavior of the system is properly described. This should make it possible to reliably derive the boost factor and to quantify the systematic uncertainties.

\subsubsection{Disc displacement system}
The disc positioning system is a complex mechanical construction. To identify the least risky and most reliable design, a study has been commissioned with an external company specialized in cryogenic engineering. Various proposals were evaluated and compared. A feasible solution has been identified, which is going to be implemented  into a technical design in the next three months.  
One of the biggest challenges remaining are the actuators used to move the discs. 
There are well-developed ideas on how the system can be constructed using piezo motors at cryogenic temperatures.

However, in order to minimize the risk, R\&D on different motors is being carried out with different companies in parallel. 
Alternative actuator crystals such as electrostrictive crystals \cite{electrostrictive} in bipolar mode are being considered as a potential solution, as they promise a larger stroke at 4~K.

Another important parameter for the disc positioning system is the time needed to adjust the disc spacing for a given frequency search. 
It is assumed presently  that rearranging the disc positions for scanning a new frequency range will not take longer than one day.
(For details on the positioning algorithm, see section~\ref{sec:pop:tuning})
This has implications on the motor design in terms of their realistic drive speed and maximum power dissipation.

\subsubsection{Disc size and tiling}
Additional main challenges are the size of the discs and the accuracy on their dimensions. 
The suitable materials have high-$\epsilon$ and low dielectric losses.
So far the best candidates are machined from Czochralski grown single crystals. For the materials under consideration, Sapphire and LaAlO$_3$, these are at best available in 12" and 3" format, respectively. This is considerably smaller than the disc diameter being aimed for. Consequently, discs from these materials will have to be tiled using gluing technology. 
While developments performed so far are promising and give some confidence that the tiling technology can  be managed with the needed accuracy for the prototype, it is not yet confirmed whether this is the case for the final experiment. Also, it is not understood how the glue gaps will influence the booster response. This issue is being investigated with simulation studies. In parallel, investigations are performed to find alternative crystals of larger size or amorphous materials.

\subsubsection{Detection technology}
The detection technology for \MADMAX{} is being investigated in two separate frequency ranges: above and below 40\,GHz. Below 40\,GHz, a receiver consisting of a low-noise HEMT preamplifier and heterodyne mixing detectors can be used to achieve the detection of signals of $\lesssim10^{-22}$\,W within a few days of measurement time. Such a receiver chain has been built and tested, and the expected performance has been demonstrated \cite{madmax_whitepaper}. 
In the near term, it is planned to implement  Josephson junction based traveling wave parametric amplifiers \cite{bib:jtwpa} to reduce the receiver noise temperature to near the quantum limit. 
The successful integration of a quantum limited amplifier into \MADMAX{} would significantly reduce the measurement time.
%This will allow to significantly reduce the measurement time, or to increase the sensitivity. 
%Institutions responsible for the development of such quantum limited amplifiers that work up to 40\,GHz have been identified, and
It is anticipated that testing 
and integration 
of such devices into the \MADMAX{} prototype can start in 2020.

The detection of signals between 40 and 100\,GHz -- the axion mass range of 120 to 400\,$\mu$eV -- remains an open question. Various possibilities such as SIS mixers \cite{sis_mixers} or single photon detectors \cite{bib:rydberg,single_photon, single_photon_kuzmin} are considered. However, a working technology will have to be developed with the help of further groups. Discussions with some groups are ongoing.  
%New technologies will have to be developed -- possibly by the cooperation between the \MADMAX~collaboration and external groups -- before \MADMAX~finishes the scan of frequencies below 40\,GHz.

\section{\MADMAX{} magnet} 
\label{sec:magnet}
A 9~T dipole magnet will be purchased for the experiment. The magnet will be delivered as complete “plug and play” solution, except for an own LHe supply. 
The overall magnet system will consist of superconducting magnet coils inside a cold mass placed in a cryostat,
a cold box for 1.8~K helium supply, a warm gas management system (except storage for gaseous helium),
power leads,
power supply and warm cabling, and the
quench protection system.

Additionally, for the shielding of the magnetic field, a suitable yoke is needed. Such a yoke \cite{yoke} is available in the DESY north hall and will be discussed in Sec.~\ref{sec:DESY}. 

\subsection{Superconducting coils}

Parametric studies have been performed to optimize the coil configuration in terms of peak field, peak forces and minimization of conductor mass.
As a result, a block design based on coils made from NbTi superconducting cable on a copper conduit was chosen. Currently, a design with 9 blocks is pursued  as a baseline.

\begin{figure}[h]
\centering 
\includegraphics[width=0.6\linewidth]{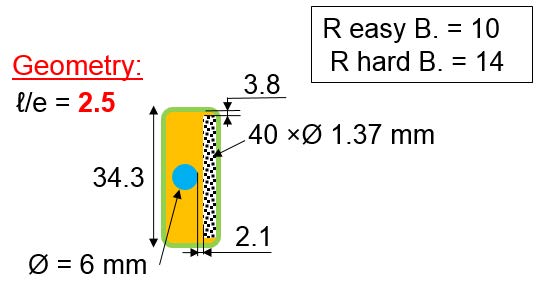}
\caption{Dimensions of the conductor for the coils of the \MADMAX{} magnet. The 40-strands Rutherford cable is shown as a dotted inlet inside the copper profile. The 6\,mm diameter hole serves for cooling the superconductor using superfluid liquid helium.}
\label{Fig:picture_cooling_concept_cable}
\end{figure}

The superconductor cable is going to be produced from a NbTi-Rutherford cable with up to 40 strands soldered onto a copper profile. 
To ensure cooling of the Rutherford cable the copper profile has a 6\,mm hole, acting as a cooling channel for the superfluid helium. The dimensions of the proposed conductor are shown in Fig.~\ref{Fig:picture_cooling_concept_cable}. A conductor with a similar design has successfully been developed for the LNCMI hybrid magnet~\cite{lncmi_magnet}. In the framework of that project a machine to reliably solder the Rutherford cable onto the copper conduit has been developed, tested and used to produce the conductor for the LNCMI magnet. This machine could be  available to the \MADMAX{} collaboration.

\begin{figure}[h]
\centering 
\includegraphics[width=0.7\linewidth]{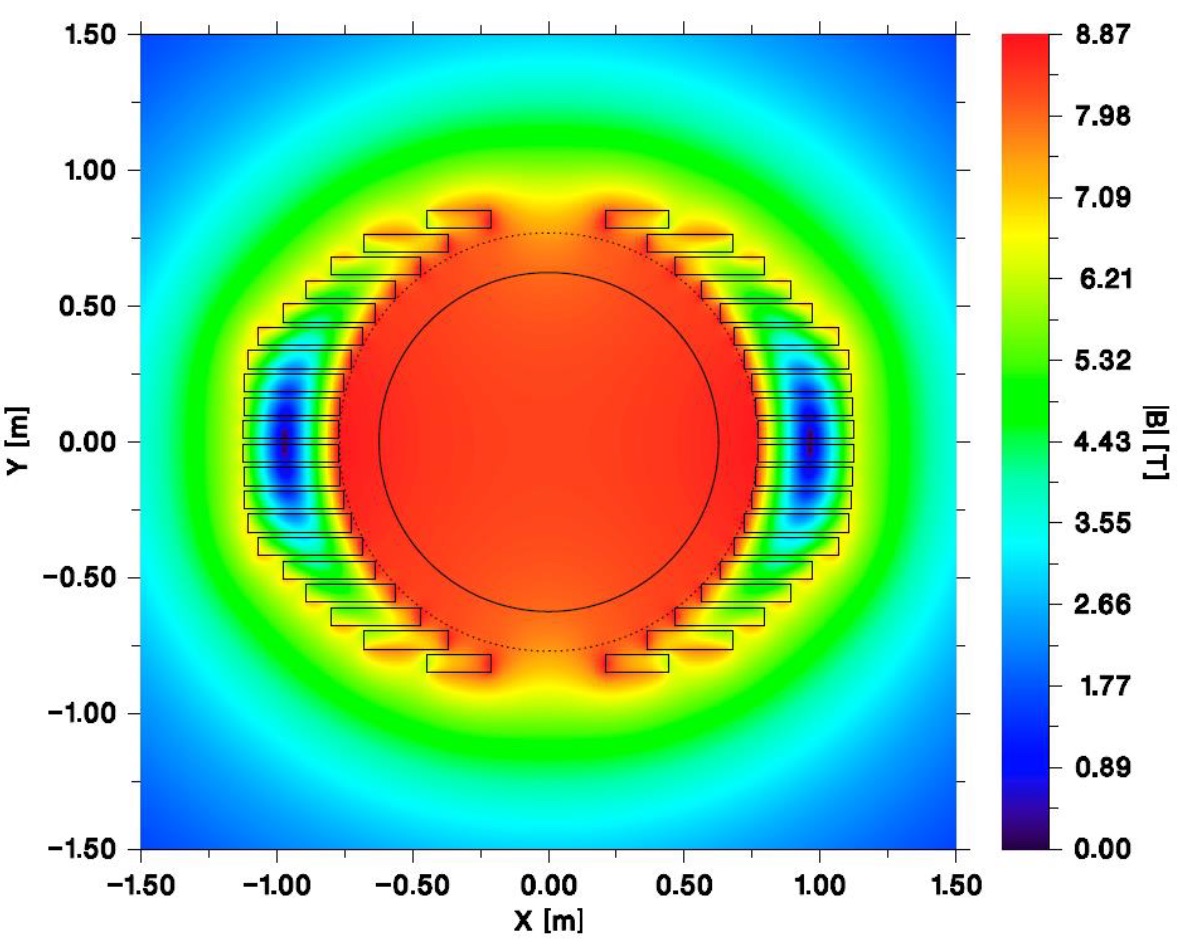}
\caption{Field configuration expected inside the 9-coils block design \MADMAX{} magnet.}
\label{Fig:MADMAX_magnet_coils}
\end{figure}

It has been shown by B-field and force calculations with finite element simulations that the conductor as specified above could withstand the forces appearing during cool-down, ramp-up and operation of the magnet. %\cite{magnet_review_report_cea}.
The resulting B-field configuration (component parallel to disc surfaces) inside and around the magnet is shown in Fig.~\ref{Fig:MADMAX_magnet_coils}.

Experience gained from the coils produced for the LNCMI magnet leads to the conclusion that the winding of the coils is feasible, using tooling available from the earlier project. Nevertheless, the first qualification plans for coil winding are presently ongoing using first dummy conductors. First results of these tests are expected in the near future. 

Given the current design, the operation current of the magnet is 23.5~kA. The connection of the power supply to the magnet will be done with normal conducting water-cooled copper cables. Inside the cryostat there will be a high temperature superconductor cable connected to the coils.

The maximum voltage during a quench should not exceed $\pm$1~kV. This is the case for the conductor itself, all electrical joints inside the cold mass and the voltage over the power leads.

\begin{figure}[h]
\centering 
\includegraphics[width=0.8\linewidth]{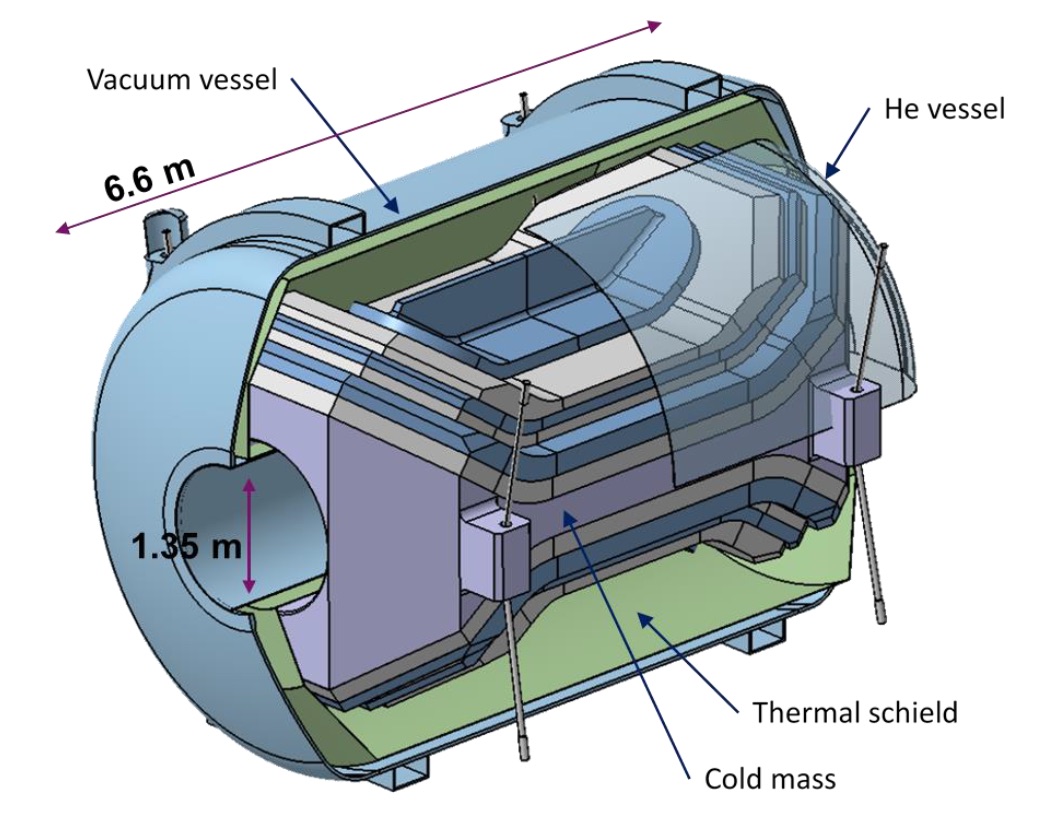}
\caption{Drawing of the superconducting coils inside their cold mass placed inside the cryostat.}
\label{Fig:magnet_sketch}
\end{figure}

\subsection{Cold mass and cooling concept}
A first design of the cold mass has been evaluated by CEA. The feasibilty of producing individual parts of the cold mass with the required tolerances is presently being investigated by Noell. The integration sequence of the cold mass into the custom made cryostat has already been studied in some detail. A sketch of the magnet structure inside the cryostat system is shown in Fig.~\ref{Fig:magnet_sketch}.

The magnet cold box and the cooling concept is part of the general order. \MADMAX{} will  provide a 4.5\,K LHe supply as described in Sec.~\ref{sec:DESY}. With a peak  field above 10\,T, the LHe for the the NbTi superconductor must be sub-cooled to achieve optimal performance. To operate the magnet on a suitable point at the load line of the superconducting cable, the nominal operating temperature of the magnet was chosen to be 1.8~K (superfluid helium). This is realized by “direct cooling”, where the coils and its structure are surrounded by the helium bath at a pressure of 1~bar. 
The concept for supplying pressurized 1.8~K helium to the magnet would be to use 3 complementary baths (B):
\begin{itemize}
\item[B1:] 4.2~K~/1~bar~$\rightarrow$ connected to the cryo-plant high-pressure supply and low-pressure return,
\item[B2:] 1.8~K/20~mbar~$\rightarrow$ saturated superfluid helium connected to a pump at 16~mbar, 
\item[B3:] 1.8~K/1~bar~$\rightarrow$  feeding the magnet bath through a helium heat pipe.
\end{itemize}
%The 1.8~K helium is necessary because of the superconductor (loadline). 
To extract the heat from the magnet and make it returnable to the cryogenic plant an easy classical heat pipe with a diameter of around 100~mm could be used due to no need of space limitation. 

The magnet heat load on the cold mass is given at this design state with less than 200\,W on the cold mass and ~3.6~kW on the 80~K shield circuit. Additionally, there will be a gas return from the power leads (HTS) on the 80~K line and the helium gas return from the compressor used for reducing the bath temperature %(bath 2) 
to 1.8\,K on the 4.5\,K line. 

\section{DESY-HH as site: Infrastructure and cryogenics}
\label{sec:DESY}

DESY has recently published the strategy document ``DESY 2030''. It includes a program for the realization of  novel particle physics experiments on the Hamburg site. Here axion and axion-like particle searches play a major role. \MADMAX{} is explicitly mentioned as part of the 
{\it "very attractive on-site programme from around 2020 onwards"}~\cite{DESY2030}.

The first axion-related experiment under construction within this program is ALPS II, aiming for first data taking in 2021. To enable \MADMAX{} and other cryogenic experiments running in parallel to ALPS II, the DESY management has decided to refurbish the cryogenic supply of the DESY north hall. Dedicated funding by the Helmholtz Association in support of the “Quantum Universe” excellence cluster of Hamburg University,
was obtained to realize this “cryogenic platform”. It shall be available by the end of 2022.

The technical requirements of \MADMAX{} on the experimental site are wide ranging. During the installation phase, heavy loads up to 200~t have to be handled, control rooms have to be provided, new transfer lines and valve boxes for the cryogenic supply of the big superconducting dipole magnet and at the 4\,K  booster vessel have to be built, and vacuum pumping systems have to be installed. At a later stage during operation and maintenance phases, most of this equipment has to stay available while challenging conditions, like a high fringe field of the magnet, have to be taken into account. Therefore, the former site of the H1 experiment at the DESY north hall seems to be an ideal place. Crane infrastructure and a hovercraft (both up to 80 t) are available for transportation and could be used during the installation phase. Electrical power, pressurized air, and cryogenic ports are also available as well as a container close to the experimental area, which could be used as a control room and electric hut. The inner space of the still existing iron yoke, used in the past for the H1 solenoid, is a very suitable place for the magnet. Supports for the magnets are available inside the yoke. First simulations demonstrated that a reduction of the magnetic fringe field down to a value of 5~mT is possible, which enables working inside the hall during magnet operation. 
The shells of the yoke can be opened to access the experimental area and closed for operation within 2 hours. Ports in the yoke allow the connection of pumping units, cabling, and transfer lines to the magnet. Space for the electrical power supply are available on a different level (6th floor) and a connection to the magnet via existing cable shafts are possible. 

\begin{figure}[h]
\centering 
\includegraphics[width=0.7\linewidth]{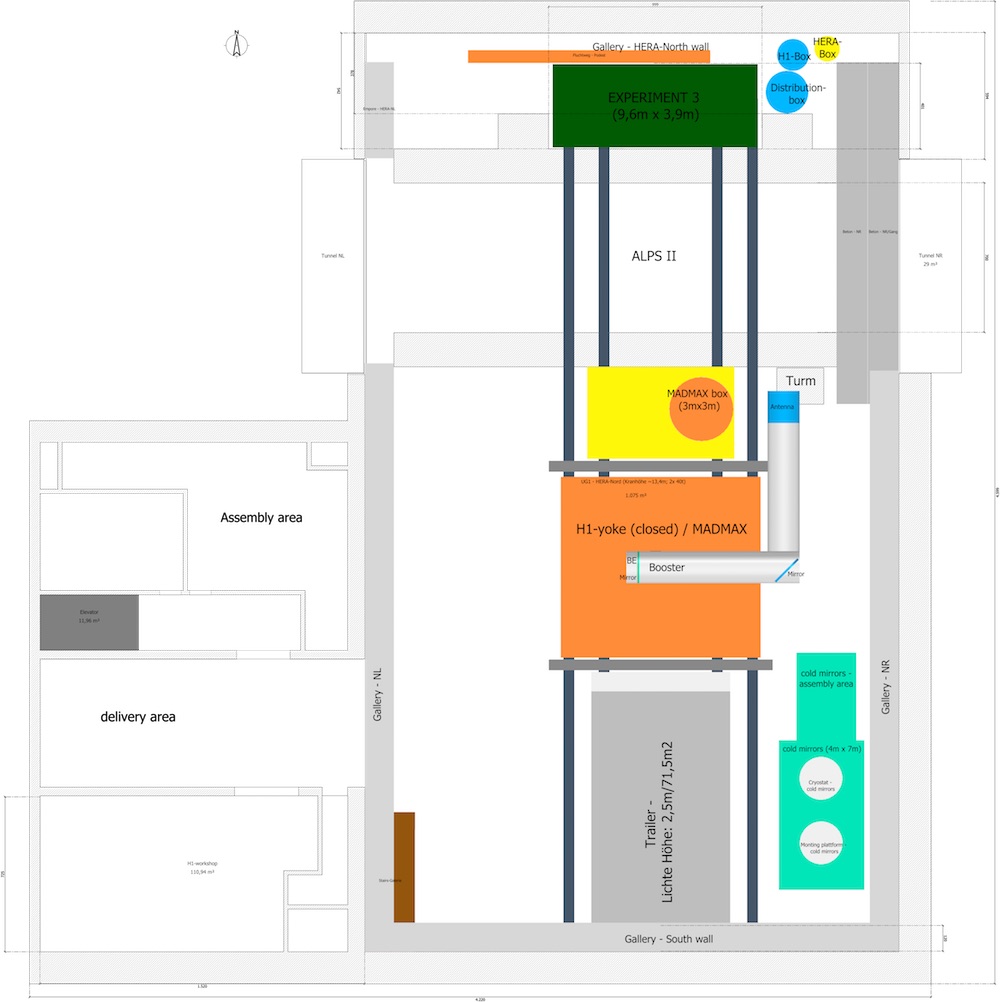}
\caption{The experimental area in the DESY north hall. The existing H1 iron yoke is in orange. The planned location of the L-shaped \MADMAX{} booster cryostat inside the yoke is indicated.}
\label{Fig:HERA_Nord_Kryoplattform_Ver3_bemasst_mitBooster_190226}
\end{figure}

For the installation of the magnet ($\approx$\,200 t) a dedicated movable crane and a hovercraft system will be rented. An overview of the DESY north hall with a place holder for the \MADMAX{} experiment is given in Fig.~\ref{Fig:HERA_Nord_Kryoplattform_Ver3_bemasst_mitBooster_190226}.

The availability of cryogenic ports for supplying liquid helium with a temperature of $\approx$\,4.2~K and also recovery lines for helium at temperatures 4.2~K and 300~K are crucial. The LHe will be provided by one of the old HERA cold boxes with a maximum power of $\approx$\,9\,kW at 4.5\,K. \MADMAX{} shares the cooling power with the other experiemtns in the hall. Investigations in cooperation with the DESY cryogenic group (MKS) shows that the nominal available cooling power is $\approx$~4\,kW at 4.5\,K. First estimates of the cooling needs for the magnet and the booster vessel indicate that this is sufficient for the cool-down and operation of the \MADMAX{} experiment. The complete warm gas management - like the 300\,K helium supply, helium gas storage and quench pipe - does also exist. All cryogenic interfaces to the DESY system have already been  identified. They are defined in Tab.~\ref{tab:Cryo_margins_HERA__North}.

\begin{table}[h]
\scriptsize
    \centering
    \makebox[\textwidth][c]{\begin{tabular}{lcccccccc} 
    \toprule
     & $T_{\rm supply}$ & $T_{\rm return}$ & $P_{\rm supply}$  & $P_{\rm return}$   &  $\Delta p_{\rm max}$  & $P_{\rm max}$    &  $f_{\rm max}$ &  \\ 
    cryo line & [K] & [K] & [\si{\bar}] &  [\si{\bar}] &   [\si{\bar}] &   [\si{\bar}]  &   [\si{\gram\per\second}] & cycles \\ 
    \midrule 
    \begin{tabular}{@{}l@{}} main \\ \SI{4}{\kelvin}\end{tabular} & \num{4.7} & \num{4.3} & \num{4} & \num{1} & \num{3} & \num{20}& \num{8} & \begin{tabular}{@{}c@{}} \scriptsize closed cycle \\   \scriptsize with connection \\ \scriptsize to quench line \end{tabular} \\  
    \begin{tabular}{@{}l@{}} shield \\ \SI{40}{\kelvin}/\SI{80}{\kelvin}\end{tabular}& $40 - 80$ & $58-66$ & $13-17$ & $12 - 16.6$ & $1 ~(\ll 1)$ & \num{20}& \num{1} & \begin{tabular}{@{}c@{}} \scriptsize closed cycle \\ \scriptsize with connection \\ \scriptsize to quench line \end{tabular} \\  
    \begin{tabular}{@{}l@{}} warm gas \\ \SI{300}{\kelvin}\end{tabular}& \num{300} & \num{300} & $16-18$ & $1$ & $15-17$ & \num{20}&  & \begin{tabular}{@{}c@{}} \scriptsize gas return \\  \scriptsize in quench line \end{tabular} \\  
    \begin{tabular}{@{}l@{}} atmosphere\end{tabular}& x & x & x & x & $20$ & &  & \begin{tabular}{@{}c@{}} \scriptsize line to atmo-\\ \scriptsize sphere exists \end{tabular} \\   
    \begin{tabular}{@{}l@{}} \SI{2}{\kelvin} (warm gas \\ return-pumping) \end{tabular}& \num{4.7} & $300$ & x & x & x & \num{20}  & \begin{tabular}{@{}c@{}}$2-5$ \\ @ \SI{15}{\milli\bar} \end{tabular} & \begin{tabular}{@{}c@{}} \scriptsize not available \\ \scriptsize to be defined \end{tabular} \\  
    \begin{tabular}{@{}l@{}} current leads \\ \SI{40}{\kelvin}\end{tabular}& $40 - 80$ & $300$ & $13-17$ & $12 - 16.6$ & x & \num{20}& \num{2} & \begin{tabular}{@{}c@{}} \scriptsize \SI{40}{\kelvin} warm \\ \scriptsize gas return \end{tabular} \\
    \bottomrule
    \end{tabular}}
    \caption{Cryogenic interfaces in DESY north hall.}
    \label{tab:Cryo_margins_HERA__North}
\end{table}

%EG %\section{Towards verification of the principle}
\section{Towards a principle validation}
It is very important to understand the systematic uncertainties on the boost factor, and its correlation to measurable quantities which can be used for tuning the positions of the discs.
In this chapter, we present simulations and an experimental setup taking realistic boundary conditions into account, such as magnetic field in-homogeneity, mechanical inaccuracies, imprecise disc geometries, etc..
In Sec.~\ref{sec:sim} we directly survey the sensitivity of the boost factor to these effects.
In Sec.~\ref{sec:meas-setup} we present a first proof of principle setup, experimentally demonstrating the tuning procedure on a small-scale dielectric haloscope, including systematic uncertainties from mechanical inaccuracies and antenna reflections.
The understanding of the boost factor gained from these studies ultimately influences the design of the prototype and of the \MADMAX{} experiment.

\subsection{Simulations}
\label{sec:sim}

\subsubsection{Methods \label{sec:sim_methods}}

Studies with a one-dimensional (1D) simulation tool are reported in~\cite{madmax_foundations}.  
To extend them to three-dimensions (3D) the following methods are applied:

\begin{itemize}
\item \textit{Finite Element Methods} ~ (FEM) can solve the full system of axion-Maxwell equations governing the \MADMAX{} booster %without making simplifying assumptions on the physics
~\cite{zienkiewicz1977finite,COMSOL,elmer,Schutte-Engel:2018mfn}. However, to make this numerically efficient for multi-wavelength-scale systems, such as \MADMAX{}, restrictions like radial symmetry needs to be assumed~\cite{Knirck:2019eug}.
 
\item \textit{Iterative Fourier Propagation} ~ Assuming zero net charges, electromagnetic waves can be transformed into momentum space with a Fourier transformation. Diffraction on the discs is applied in position space while propagation is done in momentum space. Starting from the emitted fields at each interface, the fields are iteratively propagated between interfaces until the out-propagating fields converge~\cite{Knirck:2019eug}. Using this method a significant reduction of computational effort can be achieved.

\end{itemize}

The validation of these methods for a prototype setup with  20 discs with a diameter of \SI{30}{\centi\meter} and a setup with 80 discs of \SI{1}{\metre} diameter is discussed in the following section. %The uncertainties of using an ideal three dimensional (3D) system are illustrated.

\subsubsection{Idealized booster in 3D}
\begin{figure}
    \centering
    \includegraphics[width=0.9\textwidth]{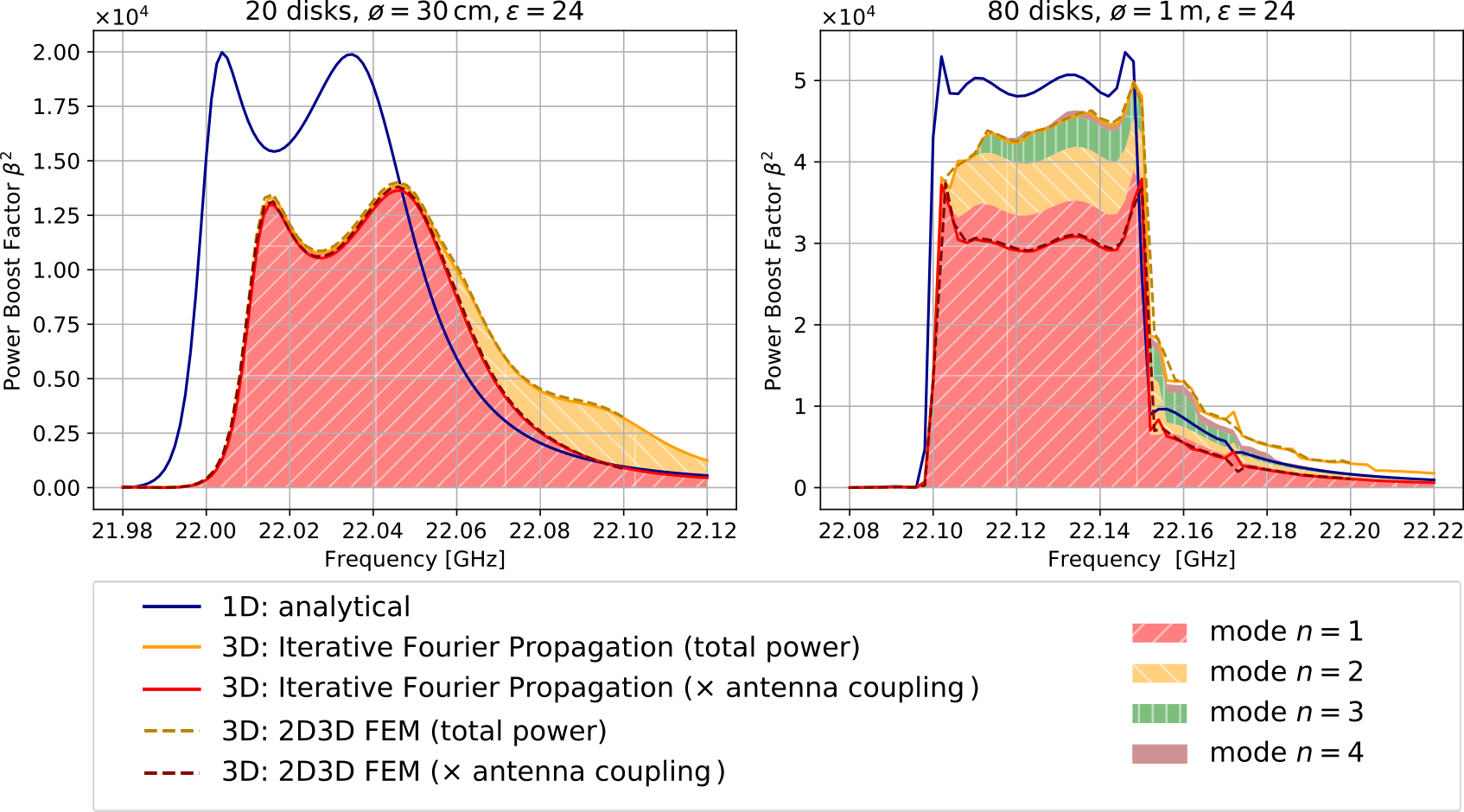}
    \caption{Boost factor considering the finite size of the dielectric discs, tuned to cover a bandwidth of $\approx$\SI{50}{\mega\hertz} at a frequency of  $\approx$\SI{22}{\giga\hertz}. Shown are the 1D analytical calculation following~\cite{madmax_foundations} (blue line), the 3D total power (orange line) and the power coupled to a Gaussian beam (red line). FEM results (dashed lines), Iterative Fourier Propagation (straight lines) and the contribution from different modes (different colored hatched areas) agree within the percent level, such that the corresponding curves lay on top of each other. \textbf{Left:} A system with 20 discs with a diameter of \SI{30}{\centi\meter} as in the proposed \MADMAX{} prototype (antenna $w_0\sim\SI{10}{\centi\meter}$). \textbf{Right:}  80 discs with a diameter of \SI{100}{\centi\meter} as in the \MADMAX{} final experiment, (antenna $w_0\sim\SI{30}{\centi\meter}$).
    }
    \label{fig:sim:benchmarkbfs}
\end{figure}

\begin{figure}
    \centering
    \includegraphics[width=\textwidth]{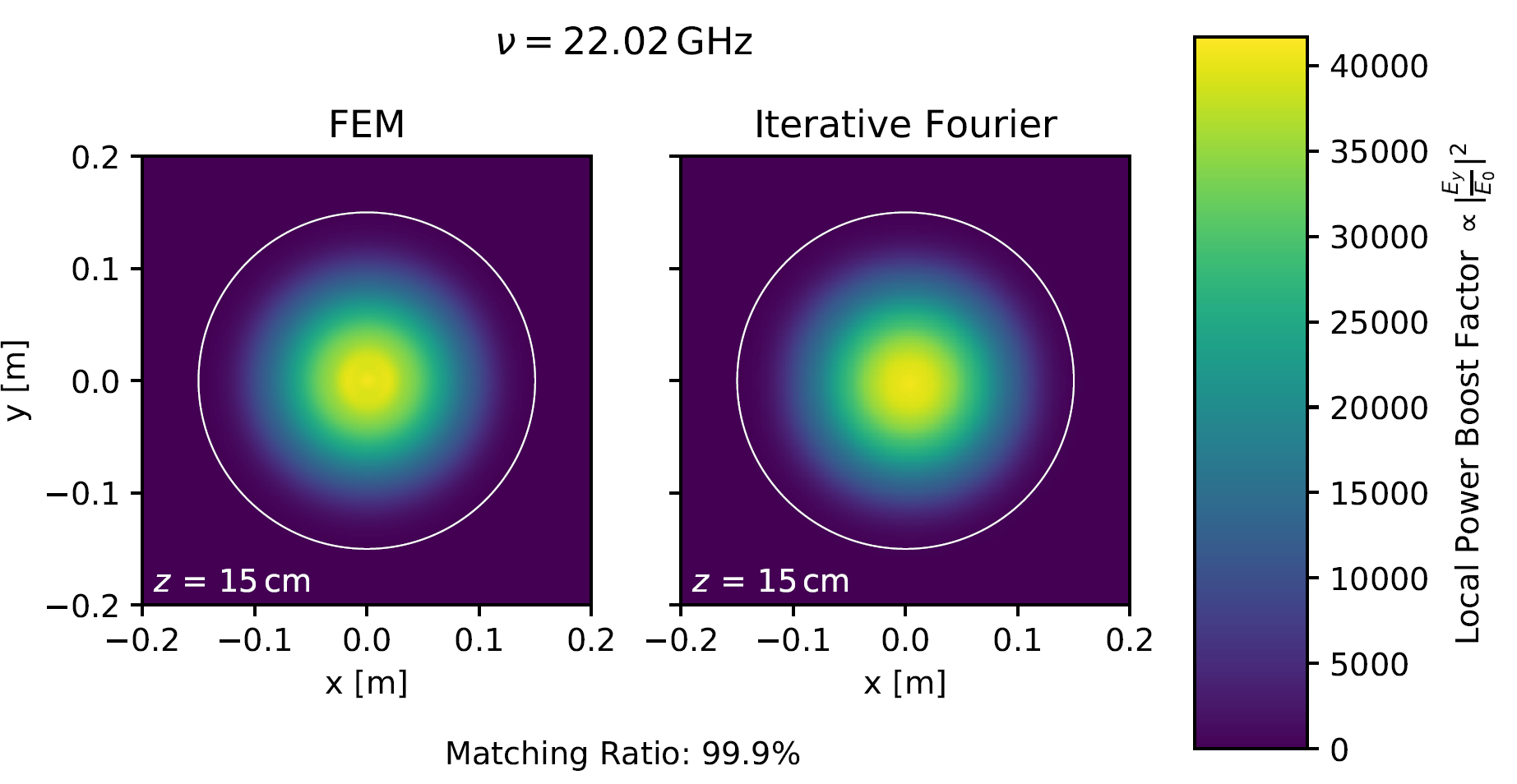}
    \caption{Comparison of beam shapes \SI{15}{\centi\metre} in front of the booster predicted by two numerical methods for the benchmark boost factor for the idealized prototype setup as in Fig.~\ref{fig:sim:benchmarkbfs} at \SI{22.02}{\giga\hertz}. The matching ratio quantifies the amount of power received by an antenna from one beam shape, if it is perfectly matched to the alternative beam shape.}
    \label{fig:sim:beamshapecomp}
\end{figure}

A comparison of the methods introduced in Sec.~\ref{sec:sim_methods}
is shown in Fig.~\ref{fig:sim:benchmarkbfs} for a booster with perfectly flat and parallel dielectric discs of finite size.
The discs are aligned such that one achieves a large boost factor over a $\sim~\SI{50}{\mega\hertz}$ bandwidth, which is our benchmark for this section.
The total power emitted by the booster is shown by the solid and dashed lines. 
Both 3D approaches give almost identical results within a few percent. 
Results from mode decomposition are shown as solid, hatched regions. The first mode (red) contains most of the axion induced power. The fourth and higher modes contribute insignificantly. %The relative contribution of higher modes is larger at frequencies away from the main boost factor peak. % resonance 
Fig.~\ref{fig:sim:beamshapecomp} shows an example for a transverse beam shape in the $x$-$y$ plane, corresponding to the frequency of \SI{22.02}{\giga\hertz} in Fig.~\ref{fig:sim:benchmarkbfs}\,(a). Here, the matching ratio between two beam shapes quantifies the amount of power received by an antenna perfectly matched to one beam, when actually the other beam is the physical beam shape.
Typically deviations between the methods are on the percent level or below, insignificant for sensitivity estimates.

Since the modes differ in transverse momentum, their respective boost factors are shifted to higher frequencies.
This can help to separate the different modes in frequency.
%, it also means that the total power emitted by a three dimensional system cannot equal that of the one dimensional calculations.
The reduction with respect to the 1D calculation is due to the specific couplings of the EM field into different modes -- analogous to the geometry factors in cavity experiments.  
%-- and not due to losses, such as diffracted power over the rims of the discs. 
%In fact, the diffraction loss is negligible for the presented configurations with perfectly flat discs.
The lowest four modes propagate almost independently in the ideal system. 
The implications deduced from the 1D calculations (see Sec.\,\ref{sec:physics_motivation}) are valid for the modes in the 3D calculation. %This including the area law 
%Apart from the individual properties of the different modes (phase velocity, loss and EM-field coupling), all implications of the 1D calculations such as the area law stated in Sec.~\ref{sec:physics_motivation} stay valid for each mode individually.

The beam shape of the lowest mode at the last disc of a booster with $\gtrsim$\,5 discs
%is given by a Bessel function of the first kind, which is zero at the rim of the discs. It 
has a 97\% matching ratio with a Gaussian beam with a waist of $w_0 \sim D/3$ where $D$ is the diameter of the discs. The power, which can be coupled to Gaussian beam antennas with such parameters, is illustrated by the red lines in Fig.~\ref{fig:sim:benchmarkbfs}. For the prototype, there is no advantage of going to a more sophisticated antenna, since for the benchmark boost factor with \SI{50}{\mega\hertz} bandwidth only the first mode contributes significantly within this bandwidth.  
The power which is not coupled into  the antenna is reflected back. The magnitude of this effect depends on the exact geometry of the focusing mirror and antenna. Such effects have already been taken into account in the first 1D calculations and compared with results from the proof of principle booster,  Sec.~\ref{sec:pop:reflections}.

\subsubsection{Requirements for an ideal booster}
\label{sec:sim:non-ideal}

The calculations described above assume zero axion velocity, a homogeneous external magnetic field, and perfectly flat, parallel, and lossless dielectric discs. In this section the systematic effects introduced by relaxing these assumptions are investigated.

As described in Sec.~\ref{sec:physics_motivation}, the axion-induced field is proportional to $\mathbf{B}_e  \, \theta$, where $\mathbf{B}_e$ is the external magnetic field and $\theta$ is the axion field. Spatial in-homogeneity of these quantities will, therefore, affect the boost factor.
However, it has been shown using Iterative Fourier Propagation
%and Mode Matching, 
that the expected magnetic field in-homogeneity, as for example shown in Fig.~\ref{Fig:MADMAX_magnet_coils}, do not affect experimental sensitivity more than to the percent level.

With non-zero axion velocity, the axion field $\theta$ acquires a spatial phase over the setup causing slightly tilted emissions from the discs~\cite{Jaeckel:2013sqa,Jaeckel:2015kea,Jaeckel:2017sjb}.
Since the axion de Broglie wavelength is much larger than the booster length, this effect is negligible. This has been explicitly verified with 1D calculations~\cite{Millar:2017eoc,Knirck:2018knd} and with Iterative Fourier Propagation.

Moreover, it is important to consider imperfections during the propagation of electromagnetic waves within the booster. 
Disc position and thickness inaccuracies cause phase errors. 
Numerical 1D calculations show that for both cases in Fig.~\ref{fig:sim:benchmarkbfs} an accuracy of $\sim \SI{5}{\micro\metre}$ is needed to limit changes of the boost factor by less than $10\%$. This dictates the accuracy of the disc positioning system.%, see also sections \ref{sec:motors} and \ref{sec:interferometer}.

%While aforementioned phase errors can be compensated by moving the discs,  
%A dielectric 
Losses modify the boost factor and limit the achievable minimal bandwidth. 
Practical implications of this are discussed in Sec.~\ref{sec:dielectrics}.

\begin{figure}
    \centering
    \includegraphics[width=0.6\textwidth]{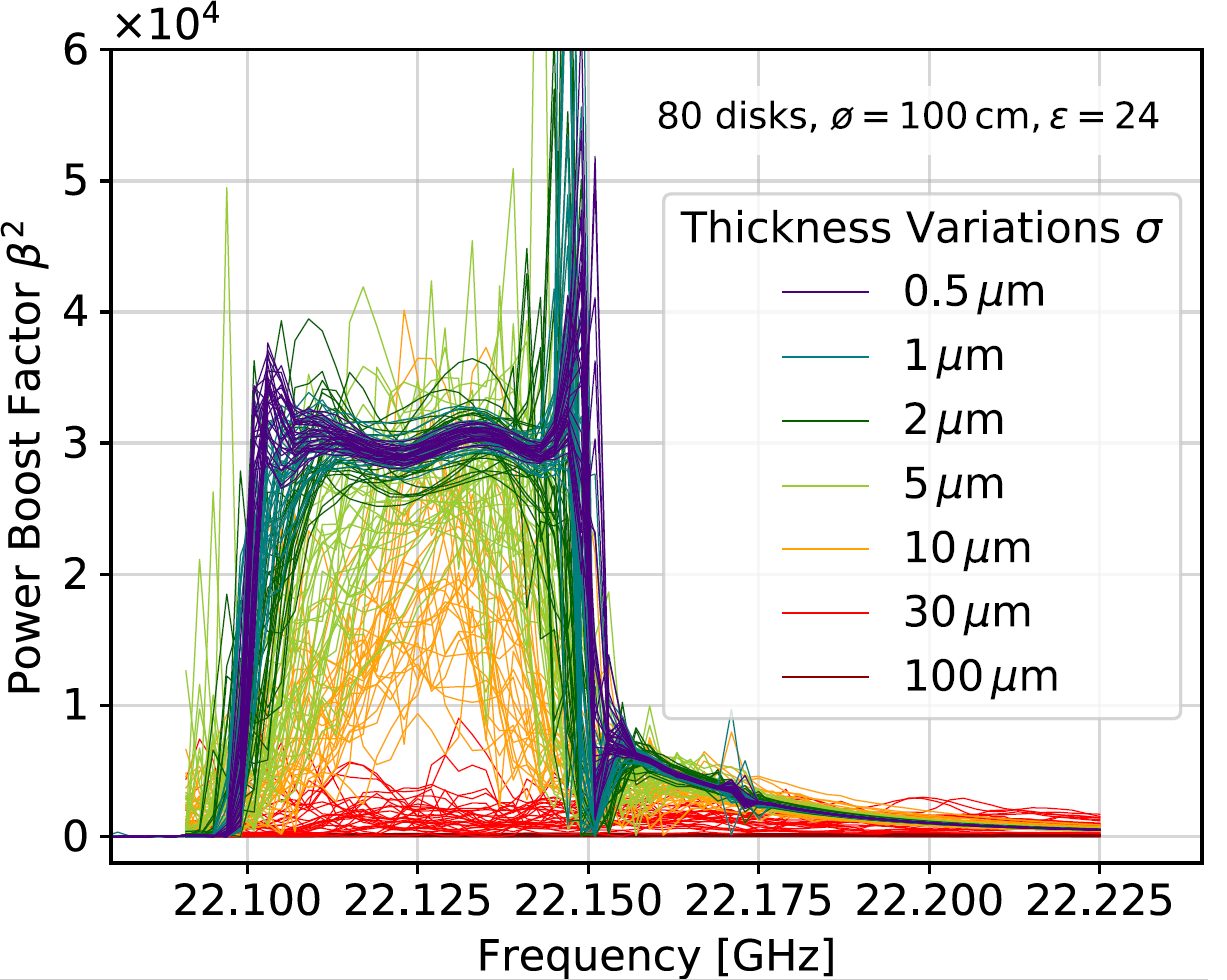}
    \caption{Effect of random disc thickness variations with a correlation length over the discs of $\sim\SI{110}{\milli\meter}$. Boost factors (including antenna coupling) are shown. Curves with different colors indicate different variance of thickness variations of individual discs. }
    %(b) Beam shape of the out-propagating wave from the booster at a frequency of \SI{22.11}{\giga\hertz} for a set of disks with random thickness variations with a standard deviation of \SI{2}{\micro\meter}. The gray dashed line indicates the contribution of the fundamental mode.}
    \label{fig:sim:surface_roughness}
\end{figure}

Thickness variations over the disc cause phase shifts. These cause slight mixing between the modes and can be directly taken into account in the Iterative Fourier Propagation approach. 
An example is shown in Fig.~\ref{fig:sim:surface_roughness} for different standard deviations of thickness variations and a correlation length of \SI{110}{\milli\meter} over the disc surface for the 80 disc booster. % This corresponds to similar length scales than the beam shapes of the first four modes.
For each thickness variation, a large set of random samples were generated, corresponding to the lines of identical color. 
 
%dark green lines, corresponding to a roughness of \SI{2}{\micro\meter} are still close to the case of an ideal disc, while the 
The light green lines correspond to thickness variations with a standard deviation of \SI{5}{\micro\meter}. They show a reduction of the area under the boost factor of the order of 30\% mainly due to the reduction of the bandwidth. 
Requiring the maximum thickness variations to be smaller than two standard deviations, indicates they should be kept \SI{<10}{\micro\meter}. 
Results for the 20 disc booster give similar requirements. 
 
In principle, using reflectivity measurements for boost factor calibration, a suitable configuration can be found re-optimizing disc positions by trial and error. At this stage, it is not clear how much the thickness variation effects can be compensated.

An analogous study was made for discs with relative tilts. % (misaligned) 
Tilts in a range from \SI{1}{\micro\radian} to \SI{3}{\milli\radian} were sampled randomly and the boost factor curve was calculated with the 3D Fourier method. The result is that  for the benchmark boost factors, tilts of \SI{0.3}{\milli\radian} for the 20 discs system, and \SI{0.1}{\milli\radian} for the 80 discs system start to affect the shape of the boost factor. % while smaller tilts have almost negligible impact. 
Note that \SI{0.1}{\milli\radian} corresponds to $\sim \SI{100}{\micro\metre}$ displacement over a 1~m diameter disc, which is less stringent than the alignment requirement. 
Additional studies are underway to understand the effect of disc tiling. Preliminary results show that the total power by the haloscope is not significantly affected by the tiling, but there can be significant changes to the modes, i.e. the beam shape.
While studies to understand the systematic effect on tiling and its possible implications on disc and antenna design are ongoing, the collaboration is also considering alternatives to tiling, as discussed in Sec.~\ref{sec:dielectrics}.

Finally, unlike cavity experiments, which are tuned to operate at high-quality factors, the boost factors of \MADMAX{} are not tuned to operate at very narrow bandwidths, but rather in a semi-broadband mode with tunable bandwidth. 
This does not only make the setup less prone to different kinds of inaccuracies, but going to higher bandwidth can relax some of the accuracy constrains stated here.
Also, the inaccuracies described here introduce mode mixing, loss, and phase errors.
The latter can be to partially compensated through a re-optimisation of the disc positions.
This is also addressed in Sec.~\ref{sec:pop:tuning}, where this possibility in a small scale experimental setup for mispositioned discs was observed. The extent to which this is possible for 20 and 80 disc setups is currently under investigation.

\subsection{Setup for proof of principle measurements}
\label{sec:meas-setup}

Stability of the boost factor during measurements necessitates that a mechanically stable booster system with the predicted electromagnetic properties can actually be built and tuned. It is therefore crucial to demonstrate experimentally that a reliable boost factor can be obtained and correlated to the available RF measurements.

In this section, first measurements with a first proof of principle booster are shown \cite{Egge:2020hyo}.
Stability of the RF behavior of the system over time and reproducibility of the boost factor have been demonstrated.

\subsubsection{Experimental setup}
\begin{figure}
	\centerline{
		\includegraphics[height=14.2em,clip,trim=130 0 420 0]{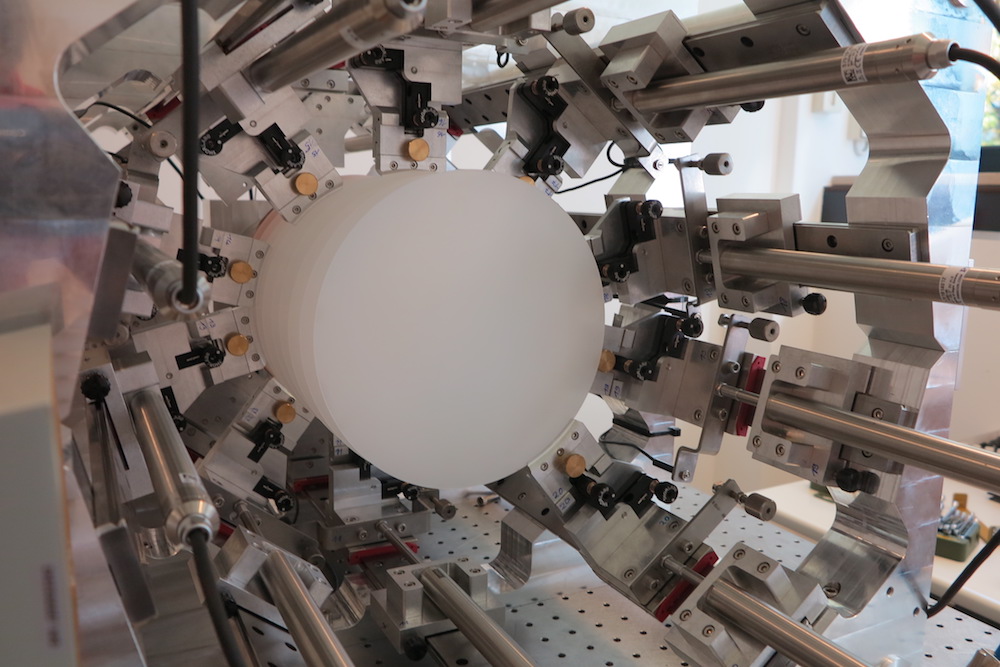}\hfill
		\includegraphics[height=14.2em,clip,trim=70 0 200 0]{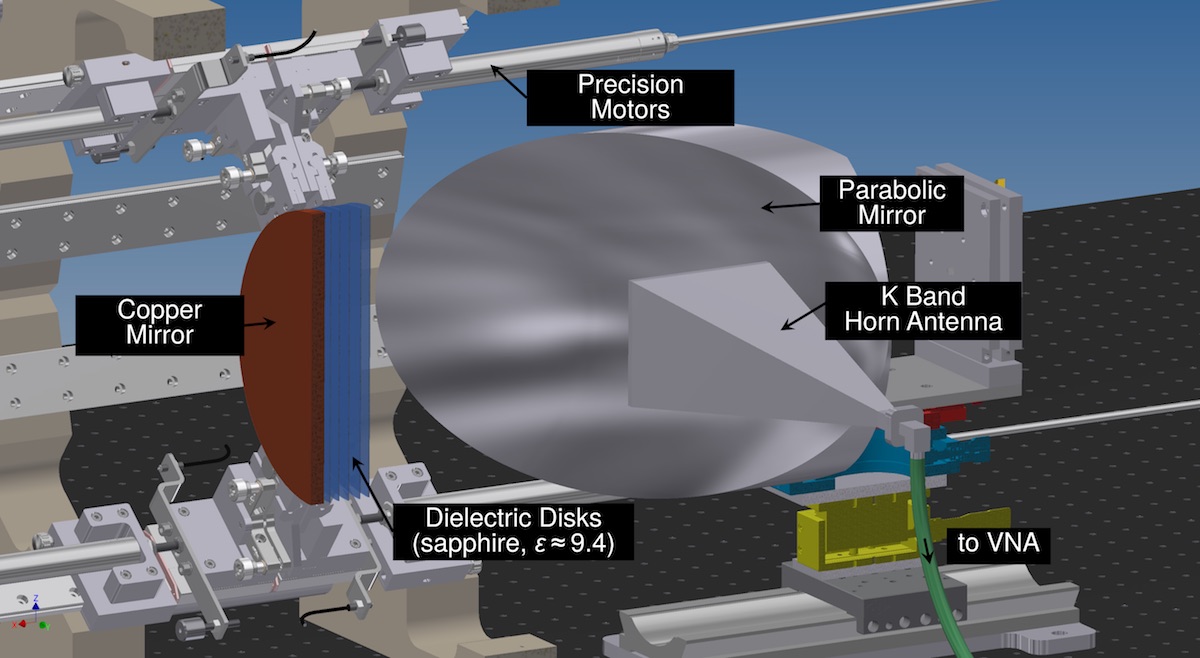}
	}
	\caption{
		\textbf{Left:} Photo of the proof of principle booster extended to 20 discs.
		\textbf{Right:} Cross-sectional drawing with 5 sapphire discs. The discs are mounted in front of a plane copper mirror and can be moved using precision motors. Via the antenna and parabolic mirror we shine in test beams and observe their reflectivity with a VNA.
		%For clarity we only show a cross section orthogonal to the disc surfaces.
	}\label{fig:pop:setup}
\end{figure}

An overview of the experimental setup is shown in Fig.~\ref{fig:pop:setup}. The proof of principle booster can carry up to 20 discs placed in front of a copper mirror.
The used sapphire discs have a dielectric constant of $\epsilon \approx 9.4$ perpendicular to the beam axis (cf. \textit{C-cut} sapphire), a thickness of $\SI{1}{\milli\metre}\pm \SI{10}{\micro\metre}$, 
and a diameter of \SI{200}{\milli\metre}. 
They are positioned by motors with a precision of less than \SI{1}{\micro\metre}.
A microwave signal can be injected into the system and the phase and amplitude of the reflected signal at a given frequency can be measured using a Vector Network Analyzer (VNA).
%(\emph{Anritsu MS4647B}). 
It is connected to a rectangular K-band horn antenna (\textit{A-INFO LB-42-25}) facing a 90-degree off-axis parabolic mirror with a diameter of $D \sim \SI{30}{cm}$.

The VNA is calibrated to the connector end of the horn antenna.
By taking a reflectivity measurement as a function of frequency without any dielectric disc installed, also the transmission
factor and the reflecitivity $\mathcal{R}_A(\nu)$ of the horn antenna and parabolic mirror assembly can be determined.

\subsubsection{Electromagnetic response model}

The electromagnetic response of the system can be described by a 1D wave-guide approximation model (1D-model). 

To experimentally study the electromagnetic properties of the booster, the reflectivity $\mathcal{R}(\nu)$ as a function of frequency is measured. It is correlated to the boost factor since the propagation of both axion-induced and reflected signals are affected by %undergo 
the same systematics. 
%Note, however, that the two types of signals might be effected differently by the different systematic effects.
%In an ideal loss-less booster the magnitude of the reflectivity will always be unity.
%

The magnitude of the measured reflectivity depends stronlgy on the loss mechanisms inside the system. Hence, these have to be understood and minimized.

It is intuitive to consider the group delay $\tau_g = - d \Phi / d \omega$, where $\Phi$ is the phase of the reflected signal and $\omega = 2\pi \nu$ is the angular frequency to extract the frequency behavior of the system. 
%For sufficiently smooth group delays it
It can be qualitatively understood as the mean retention time of reflected photons within the booster and therefore maps out resonances.
%The correlation in frequency between b
Boost factor and group delay for a set of four equally spaced discs at \SI{8}{\milli\metre} distances are compared in Fig.~\ref{fig:pop:gd-incl-reflection}. 
The group delay peak at the highest frequency correlates to the one from the boost factor from the 1D calculation.

\subsubsection{Antenna reflections}
\label{sec:pop:reflections}
\begin{figure}
	    \centering
	    \includegraphics[width=0.7\textwidth]{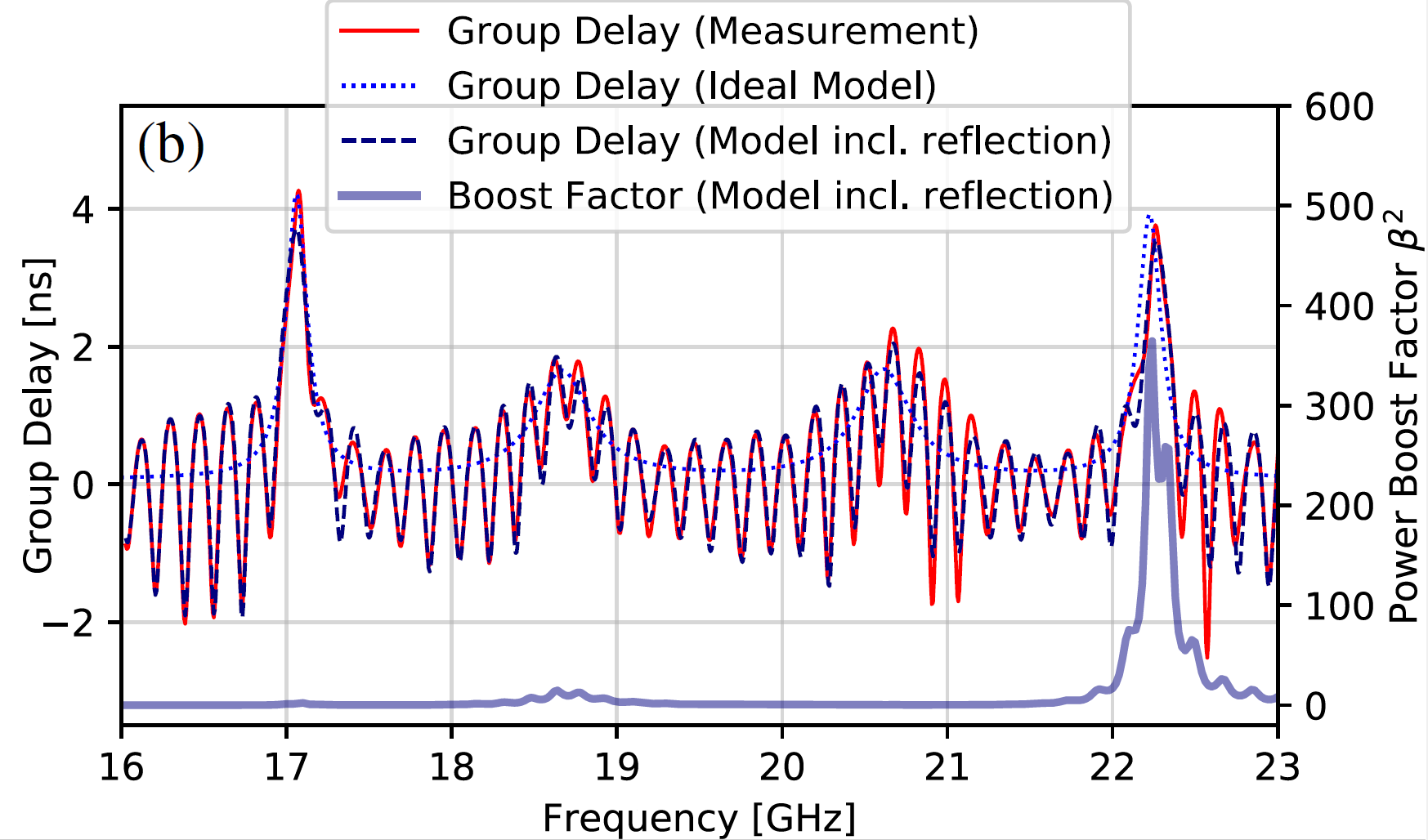}
	    \caption{Group delay including antenna reflection from measurement (red straight line), without antenna reflection in idealized loss-less model (blue dotted line) and after including the antenna reflection (dark blue dashed line). Also the boost factor including the reflection on the antenna is shown (thick light blue line).}
	    \label{fig:pop:gd-incl-reflection}
	\end{figure}

An important systematic effect on the measurement of the electromagnetic response and ultimately on the boost factor is from unwanted reflections.
Fig.~\ref{fig:pop:gd-incl-reflection} shows the effect of these reflections, illustrating the consequences to the group delay.
The reflected signal is out of phase with the main signal and can cause constructive or destructive interference, adding harmonics (wiggles) to the frequency spectra of the group delay.
It potentially deteriorates the shape of the peaks and thus can lead to an inaccurate determination of their maxima.
More importantly, the unwanted reflections cause deterioration in the boost factor itself.

	Therefore, it is important to include these reflections in the 1D-model.  This can be done by including the measured antenna reflectivity $\mathcal{R}_A(\nu)$.
    Fig.~\ref{fig:pop:gd-incl-reflection} shows the results from such a model compared with a measurement for a four disc case. The features imposed by the reflections are reliably reproduced.

\subsubsection{Mechanical stability}
The mechanical stability of the system is determined by higher frequency (seismic) vibrations from the surrounding and  seismic movements coupling into the booster system, as well as long term displacement effects during measurement caused, for example, by temperature changes. 
These will depend on the location of the setup  and the precautions taken in the design regarding thermal stability and vibration damping.

The long-term behavior of the proof of principle system was investigated by observing the stability of the resonant group delay peak of a single disc and mirror system. Measurements taken over more than a day without moving any motors revealed a temperature dependence of the group delay peak position of  $\sim~\SI{2}{\mega\hertz \per \kelvin}$ (corresponding to a dependency of the disc position of around \SI[separate-uncertainty = true]{-7(2)}{\micro \meter \per \kelvin}) \cite{egge}. 
Also considering the higher frequency vibrations it could be concluded that without significant precautions the group delay peak was stable to $\sim$\,1\,MHz, limited by the seismic constraints.

The accuracy of the disc alignment is limited by the precision of the motors and the mechanics of the disc holder. 
The reproducibility of positioning individual discs was determined by repeatedly placing a disc to a predefined position. The obtained group delay peaks were used to calculate the position by comparison to the 1D-model. Hysteresis of the system is taken into account.
The group delay peak position was reproduced within a standard deviation of $\sim \SI{1.5}{\mega\hertz}$ corresponding to position errors of the order of  $\sim \SI{700}{\nano\metre}$.
This is still well below the needed accuracy for the benchmark boost factors studied in Sec.~\ref{sec:sim}.

\subsubsection{Tuning routines}
\label{sec:pop:tuning}
\begin{figure}
    \centering
    \includegraphics[width=0.4\textwidth]{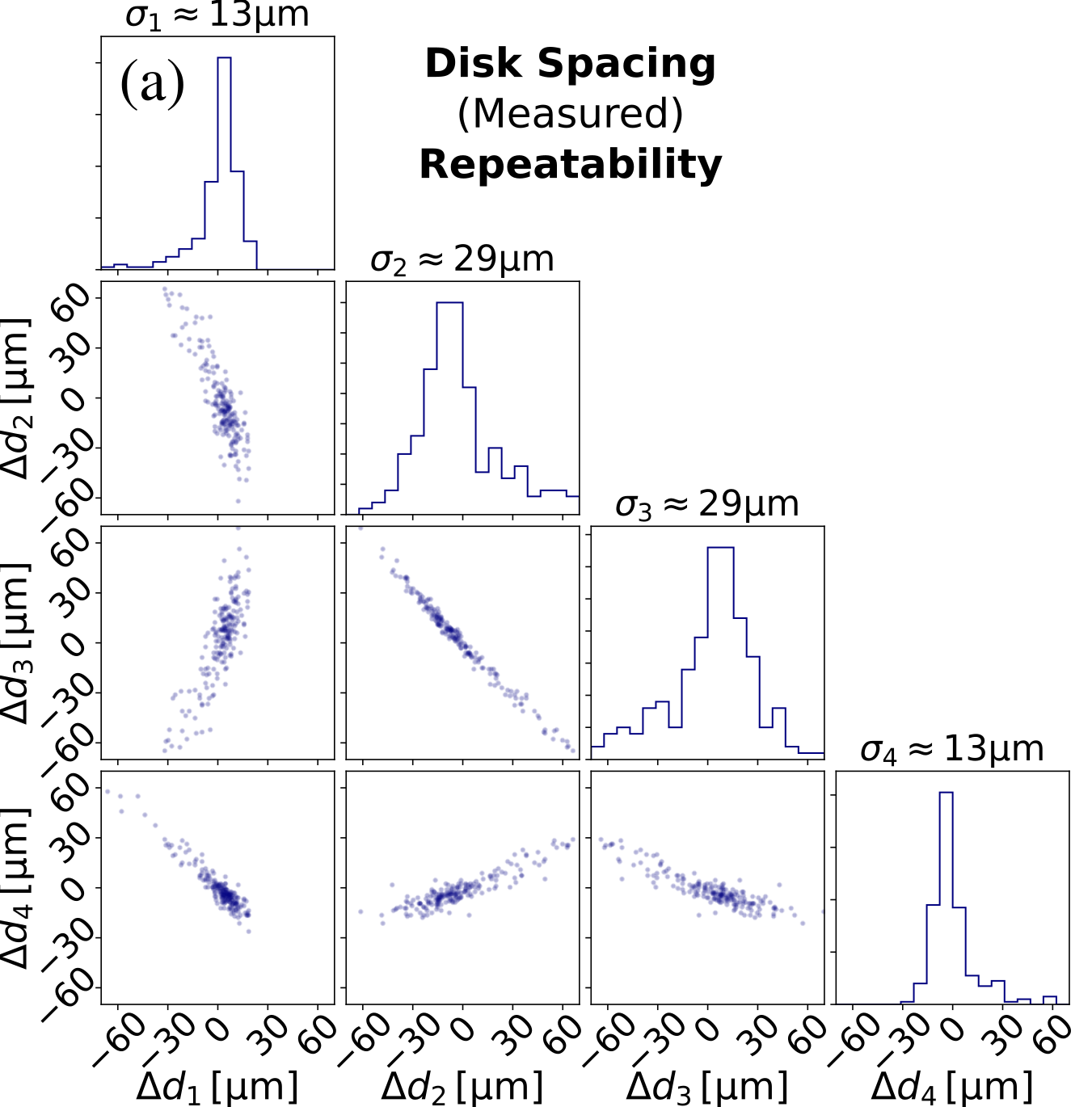}
    \includegraphics[width=0.5\textwidth]{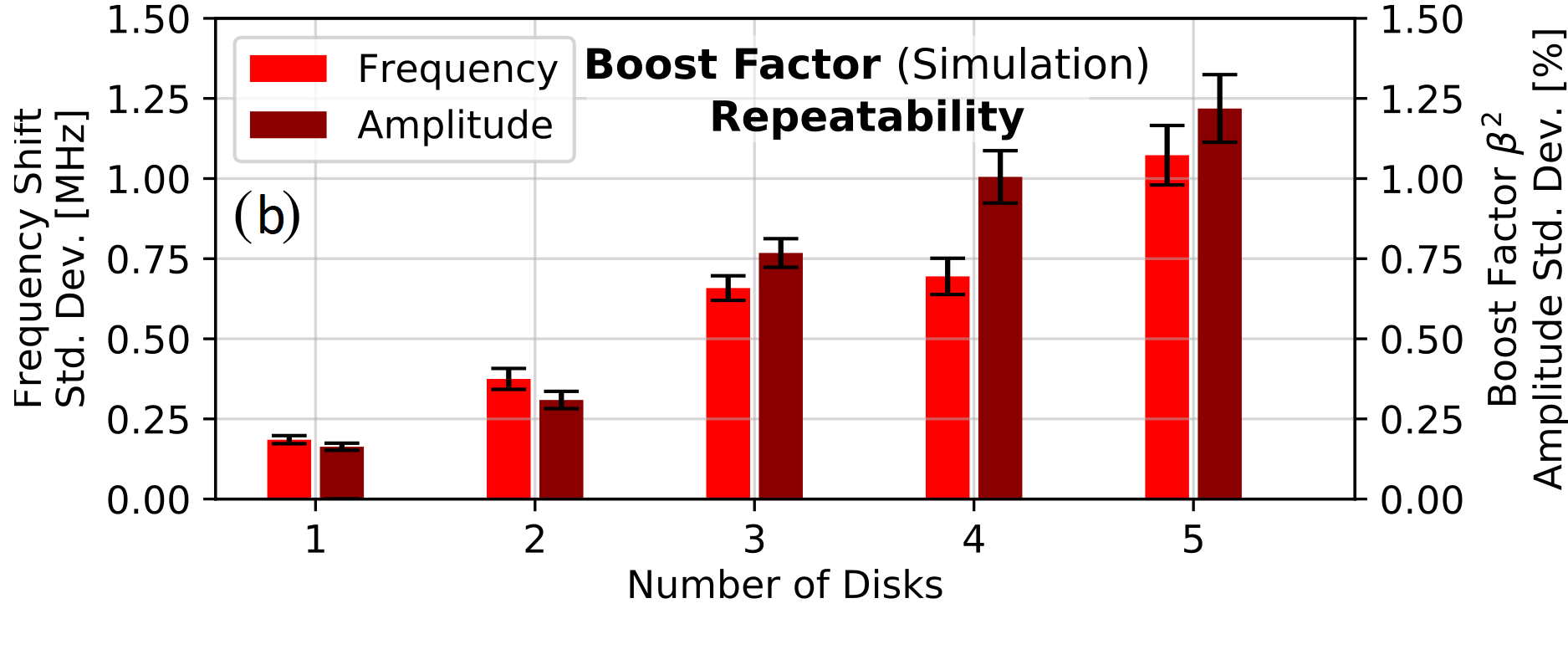}
    \caption{(a)~Correlation of disc spacings after optimizing the actual physical disc positions for the example of a 4 disc setup for $\sim 200$ times. $\Delta d_1$ refers to the spacing between the copper mirror and the first disc, $\Delta d_2$ the following spacing, and so on. (b)~Standard deviation on frequency position (left bars, light red) and standard deviation on amplitude relative to the ideal amplitude (right bars, dark red) of the boost factor after fitting the model to the measured group delays corresponding to the physical disc positions shown in (a),(b). More details cf. text.}
    \label{fig:pop:tuning}
\end{figure}

During the axion search, the boost factor function needs to be tuned in frequency. It could be demonstrated with up to five equidistant discs that such tuning can be achieved using the following tuning procedure: 
First, the desired boost factor function is calculated via the 1D-model. The same is done for the corresponding group delay. 
The physical disc positions in the setup are then adjusted by an optimization algorithm. For each optimization step, the discs are realigned, and the group delay is measured and compared with the corresponding model. Based on the obtained difference new disc positions are calculated.  
The optimization algorithm stops once the new disc positions converged within \SI{1}{\micro\metre}.
By repeatedly performing this algorithm using different starting positions within \SI{100}{\micro\metre}
the reproducibility of the system response can be evaluated.

Fig.~\ref{fig:pop:tuning}~(a) show the results of running the frequency tuning procedure for $\sim 200$ times using the same calculated boost factor function with four discs \cite{egge}: 
the final motor positions are correlated to each other. 
The visible correlations show that there exists degeneracy in the disc position phase space.

To quantify the systematic uncertainty on the boost factor arising from tuning,
the optimization algorithm was used on the measured group delays with the final disc positions by adjusting the calculated ones in the 1D simulation until they match the measured ones. Afterwards, for each realization, the corresponding boost factor was calculated. The systematic uncertainty on the obtained maximum of the boost factor can then be extracted from this ensemble, thus taking into account the effect of disc spacing degeneracy.
Fig.~\ref{fig:pop:tuning}\,(b) shows the resulting uncertainties on the boost factor center frequency and amplitude as a function of the number of discs for a measurement for setups resulting in a boost around 22\,GHz (see Fig.~\ref{fig:pop:gd-incl-reflection}. For this case, the frequency uncertainty ($< \SI{2}{\mega\hertz}$) and the amplitude uncertainty ($< \SI{2}{\percent}$) would have a small impact on the sensitivity of the booster. 
Analogous results have been obtained for other frequency range between \SI{22}{\giga\hertz} to \SI{28}{\giga\hertz}.

The correlations in disc spacings and the resulting boost factor functions illustrates that different disc spacing configurations can still result in the same electromagnetic response. This shows that
%when compensated with the other spacings.
%Here an advantage of tuning using multiple parameters is revealed: 
phase errors, such as position errors of discs, can be compensated by corresponding phase errors resulting from other discs. In principle, this degeneracy could be also utilized to compensate for the sum of phase errors resulting from disc and position imperfections.
%Therefore, accuracy constrains -- in particular also those derived in section~\ref{sec:sim:non-ideal} -- most likely be loosened considering such a tuning algorithm.
%The corresponding uncertainties on the boost factor introduced by the optimization scheme outlined in figure~\ref{fig:pop:tuning}\,(c) make this point explicit.
%Therefore, it is currently being investigated how such a reoptimization  section~\ref{sec:sim}

\subsubsection{Outlook}

\begin{figure}
    \centering
    \includegraphics[width=0.35\textwidth]{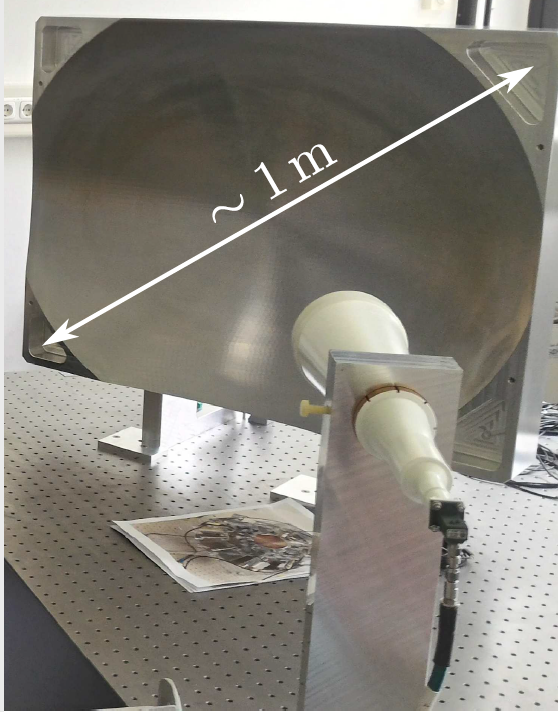}
\includegraphics[width=0.55\textwidth]{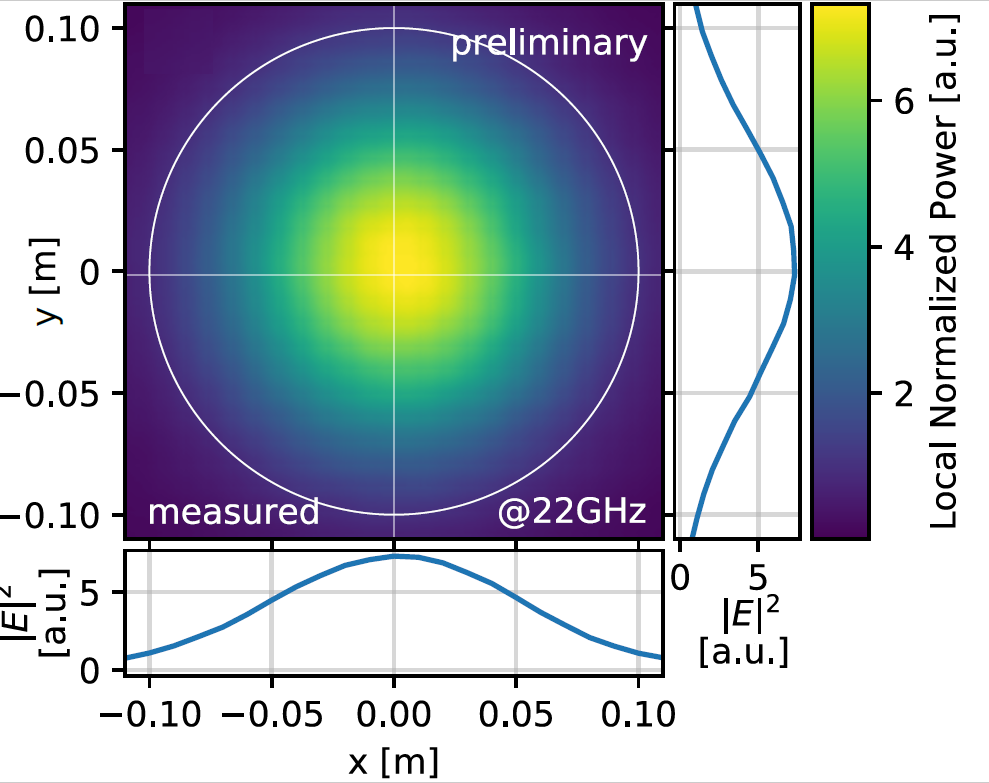}
    \caption{Left: New Gaussian horn and ellipsoidal mirror. Right: Result of beam shape measurement $\sim \SI{70}{\centi\metre}$ in front of ellipsoidal mirror. 
    }
    \label{fig:pop:new_stuff}
\end{figure}{}

The setup is currently being extended to 20 discs, as shown in Fig.~\ref{fig:pop:setup}\,(left).
%To this end we further
Antenna reflections have been reduced by employing an antenna which is better coupled to the lowest mode of the booster. A custom-made Gaussian horn antenna, together with a large ellipsoidal mirror from \textit{Thomas Keating Ltd.}
shown in Fig.~\ref{fig:pop:new_stuff}\,(left), has recently been installed, reducing the antenna reflections by $\sim \SI{10}{\decibel}$. The shape of the reflected beam at 22\,GHz has been measured using a near-field probe and is shown in Fig.~\ref{fig:pop:new_stuff}\,(right). It agrees well with a Gaussian beam.

Additionally, an optical low-coherence interferometer \textit{SOFO V}~\cite{madmax_interferometer} (also see Sec.~\ref{sec:interferometer}) will be utilized to independently measure the actual disc positions and tilts. This can help in understanding and improving 3D simulations.

%\section{Antenna system}f
\section{Detection system}
The detection system of \MADMAX\ can be divided into the optical system -- consisting of the antenna and focusing mirror -- and the receiver. 
The requirement for the optical system is a high efficiency of coupling the axion induced signal into the receiver while avoiding coupling to the thermal radiation from the surrounding. As the expected signal power is only of the order 10$^{-22}$\,W, a very low noise receiver is needed.  
%While there exist mature technologies for both, the receiver system requires further investigations in order to lower the noise temperate at frequencies $<40$ GHz and achieve the target sensitivity at $>40$ GHz.
\subsection{Optical system}
\label{sec:optical_system}
\begin{figure}[!h]
    \centering
    \includegraphics[width=0.7\textwidth]{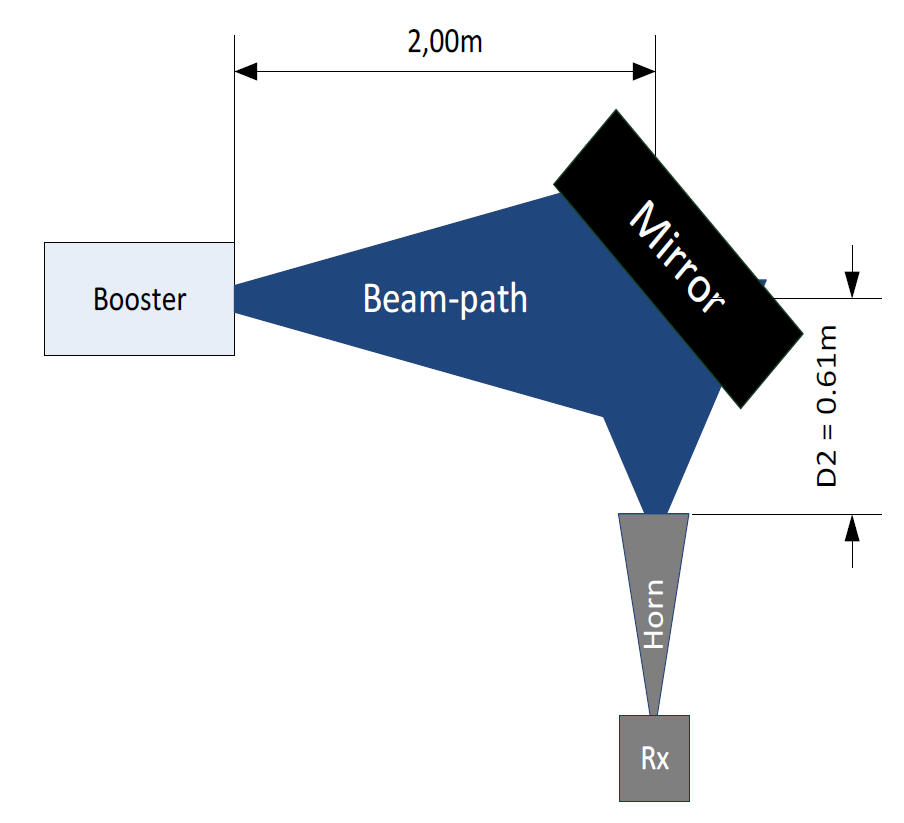}
    \caption{Sketch of the \MADMAX{} optical system collecting and guiding the signal from the prototype booster to the receiver system.}
    \label{fig:optical_system}
\end{figure}

The axion induced radiation emitted by the booster is guided and focused onto the receiver by the optical system. It consists of a focusing mirror, a horn antenna and waveguiding inside the booster cryostat. A simplified sketch of the system is shown in Fig.~\ref{fig:optical_system}.

The size of the focusing mirror for the prototype, sitting at a distance of $\sim$2\,m behind the booster, has been optimized for the ideal coupling of the Gaussian beam shape into the horn antenna. The size of the mirror is related to the expected beam waist of the booster signal. This was determined from the simulation.
This leads to a diameter of the focusing mirror of $\sim$\,1\,m, which determines the dimensions of the cryostat. Note that the shape of the focusing mirror is elliptical and it is placed inside the cryostat with an angle, this allows to place it inside a tube with an inner diameter of 750\,mm.
The horn antenna has a length of $\sim$\,350\,mm and an opening width of $\sim$\,150\,mm. It produces a nearly perfect circular beam. The first visible side lobe at design frequency is at -35\,dB. The coupling integral to the Gaussian beam (first mode) is $>95\%$ in the frequency range between 18 and 25\,GHz.     

For the final experiment the expected mirror diameter for coupling to the Gaussian mode of $\sim$95\% is extrapolated to be around 2\,m, leading to a necessary cryostat inner diameter of $\approx$\,1.4\,m. 
The final dimensions of the system will be determined once the exact beam waist for the 80 disc booster system is available.

\subsection{Detector technologies}

%\begin{itemize}
%\item Existing system: Heterodyne mixing with HEMT
%\item Plans with JPA development - TWA
%\item Long term plans for single photon detection
%\end{itemize}

The power of the axion signal from the booster is expected to be $\sim10^{-22}$ W. In order to detect such a small signal with up to a week of measurement time, a receiver with very low noise is required. At frequencies below $\sim40$ GHz, there exists a mature technology -- the High Electron Mobility Transistor (HEMT) based amplifier -- that has a noise temperature of a few Kelvin when operated at liquid helium temperature. The baseline design of the receiver chain for data taking between $10$ to $40$ GHz based on HEMT preamplifiers will be discussed in section~\ref{sec:hemt}. It has been demonstrated in the lab that the aforementioned receiver chain is capable of detecting signals of $10^{-22}$ W within a few days.

 In the intermediate term, the receiver noise temperature can be significantly reduced by replacing the HEMT amplifier with a quantum-limited amplifier, such as a SQUID or Josephson Junction based parametric amplifier. Such technologies have already been used by other axion search experiments~\cite{bib:admx,bib:haystac,bib:quax}, though they are typically operated at lower frequencies and with narrower bandwidth compared to \MADMAX. Section~\ref{sec:quantumamp} will discuss the potential of reducing the receiver noise temperature to $\sim1$ K by using quantum-limited linear amplifiers for \MADMAX~operation up to $40$ GHz. For axion detection beyond $40$ GHz, new technologies will be needed, and various possibilities are to be explored. Section~\ref{sec:highfreq} briefly describes a few possibilities that have been considered and the path forward for \MADMAX.

\subsubsection{HEMT-based receiver chain}
\label{sec:hemt}
For microwave detection between $10$ to $40$ GHz, a linear detection scheme based on heterodyne mixing is typically used, whereby the signal to noise ratio (SNR) is given by Dicke's radiometer equation,
\begin{equation}\label{eq:dicke}
\text{SNR}=\frac{P_{sig}}{k_BT_{sys}\Delta\nu\sqrt{\frac{1}{t_{scan}\Delta\nu}+\left(\frac{\Delta G}{G}\right)^2}},
\end{equation}
where $k_B$ is the Boltzmann constant, $P_{sig}$ is the expected axion signal power corrected by detection efficiency. $T_{sys}$ is the total system noise temperature which consists of the receiver noise temperature $T_{rec}$ and the additional noise from the booster and its surroundings $T_{booster}$, such that $T_{sys}=T_{rec}+T_{booster}$. $\Delta\nu=10^{-6}\nu_a$ is given by the axion line width and $t_{scan}$ is the integration time for an individual measurement of a given bandwidth. $G$ and $\Delta G$ are the receiver gain and gain fluctuation, respectively. Since the axion signal would be a narrow line in the measured power spectrum, the term arising from gain fluctuation can be neglected and thus Eq.~\ref{eq:dicke} can be reduced to
\begin{equation}\label{eq:dickesmpl}
\text{SNR}=\frac{P_{sig}}{k_B T_{sys}}\sqrt{\frac{t_{scan}}{\Delta\nu}}.
\end{equation}

\begin{figure}[!h]
\centering
\includegraphics[width=1\textwidth]{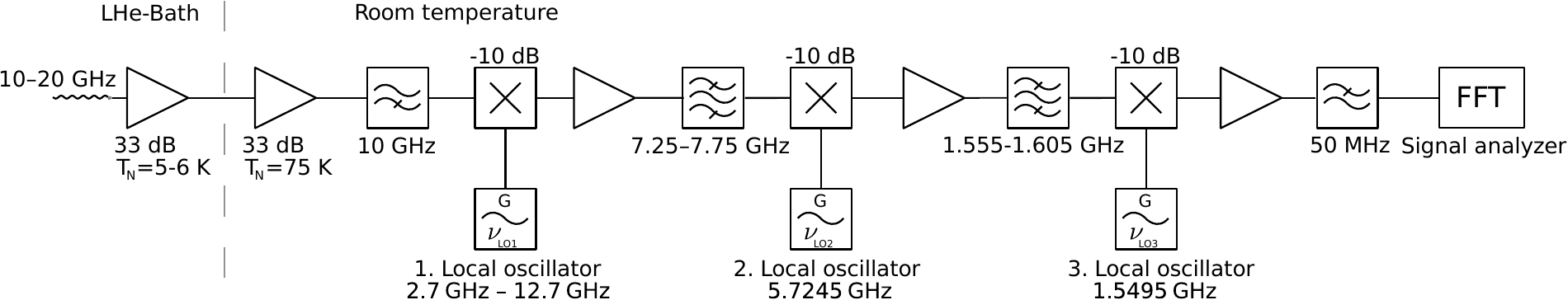}
\caption{Diagram of the receiver chain.}
\label{fg:receiver}
\end{figure}

\begin{figure}[!h]
\centering
\includegraphics[width=0.5\textwidth]{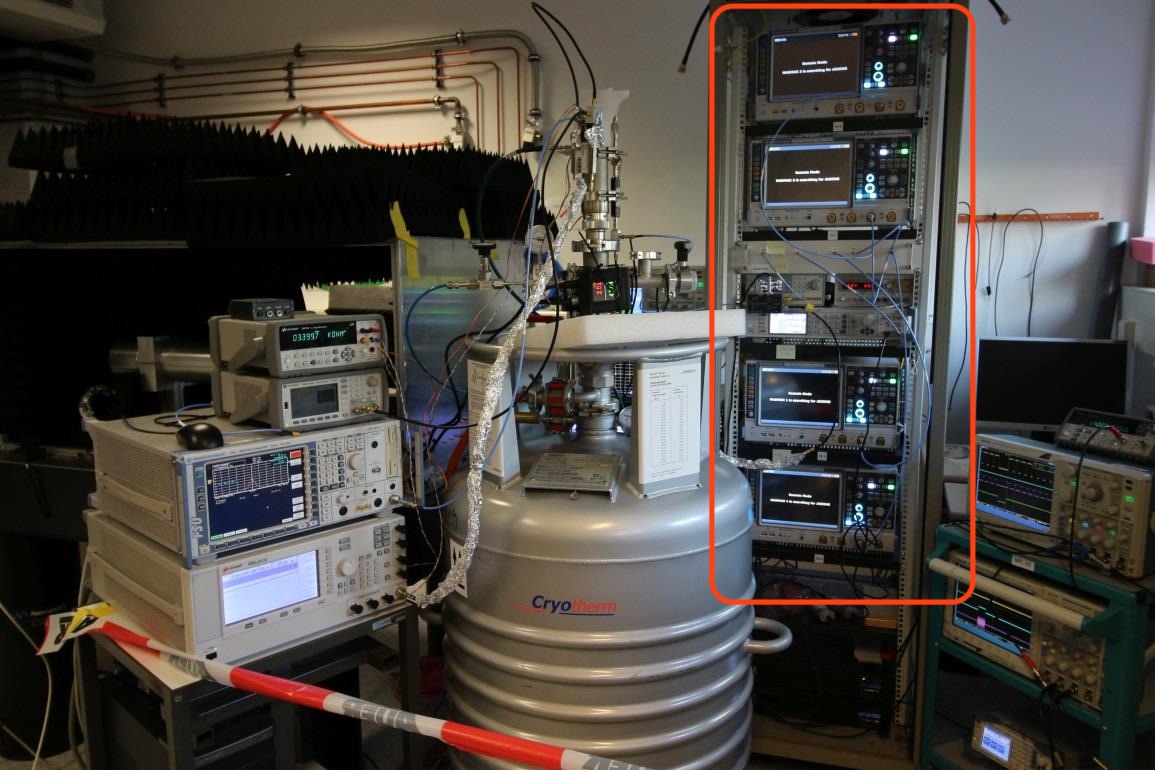}
\caption{The digital 16-bit samplers with internal FPGAs (indicated by the red box) that provide fast Fourier transform. The LHe dewar in which the HEMT was operated and the oscillators that produced the fake axion signal are also visible.}
\label{fg:samplers}
\end{figure}

%Figure~\ref{fg:heterodyne} shows a diagram of heterodyne detection. The signal is first amplified by a low noise preamplifier (labeled RF), and then shifted to intermediate frequencies (IF) by the mixing with a carrier frequency. At intermediate frequencies the signal is further amplified and filtered. 
%A receiver chain could consist of multiple mixing stages such that the signal is shifted to a frequency accessible to digital 16-bit samplers, which have internal FPGAs that provide real-time fast Fourier transform calculations with subsequent integration and storage of the signal.
Fig.~\ref{fg:receiver} shows the proof-of-principle receiver chain that has been setup at MPP. It consists of a low-noise cryogenic preamplifier and three-stage heterodyne receiver with subsequent fast Fourier signal analyzer. The InP-HEMT preamplifiers from Low Noise Factory can achieve an input-referred noise temperature of $5-6$~K and a gain of $33$~dB when operated at an ambient temperature of 4~K. The output of the preamplifier is sent through an amplifier at room temperature and then down converted to a center frequency of $26$~MHz with a bandwidth of $50$~MHz. The data acquisition system consists of four time-shifted digital 16-bit samplers with a sampling rate of $200$ MS/s and internal FPGAs for real-time FFT calculation and subsequent averaging. Each sampler takes turns to record the signal with $\sim20$~ms intervals in an alternating fashion while FFT calculation is being performed by the other three. The dead-time is reduced from $75\%$ with only one sampler to less than $1\%$, and yields a $2.048$~kHz bin width in the measured power spectrum. Fig.~\ref{fg:samplers} shows the samplers at MPP Munich.\\

\begin{figure}[!h]
\centering
\includegraphics[width=0.6\textwidth]{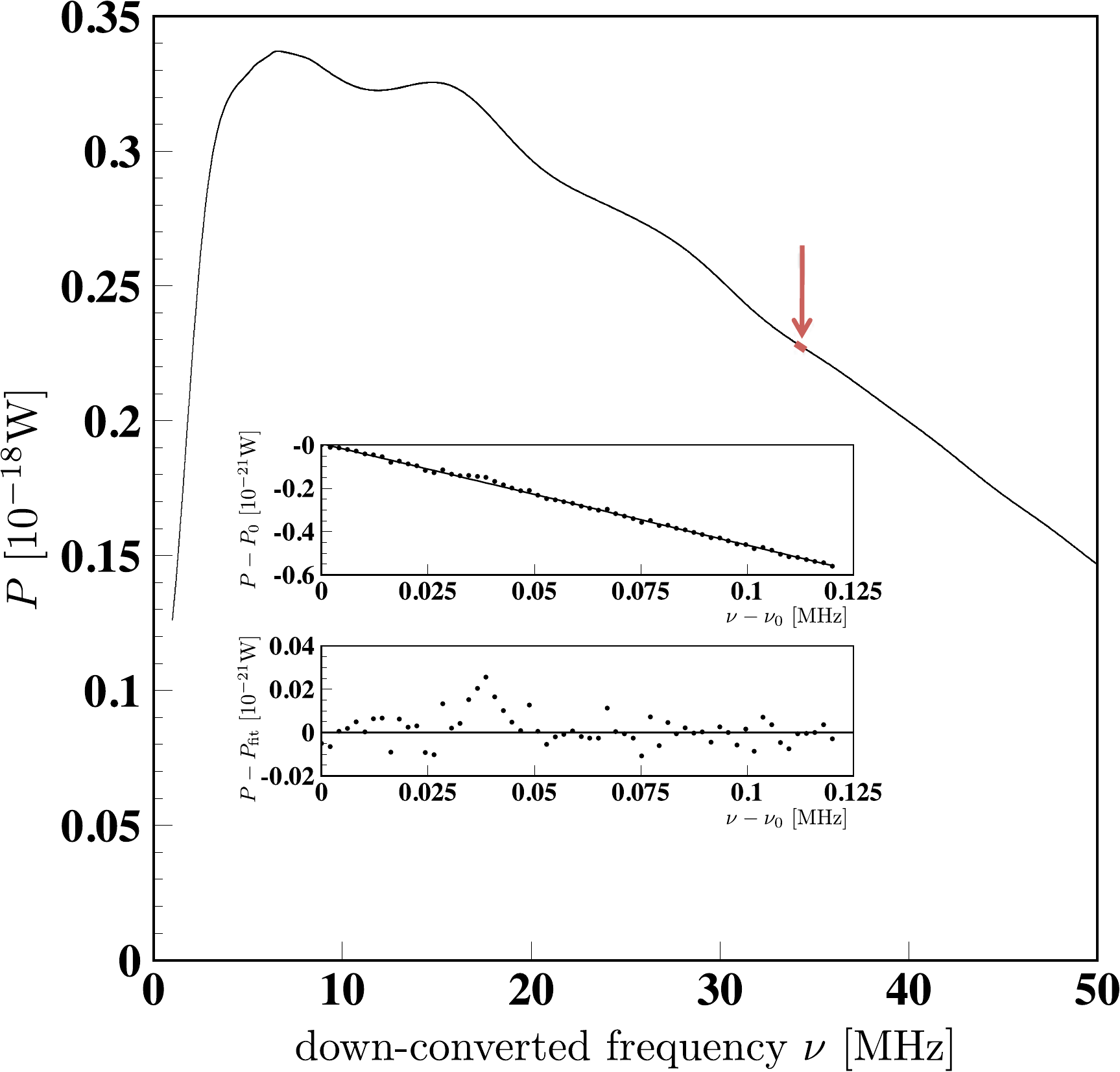}
\caption{Test signal at $18.4$ GHz with power of $\sim1.2\times10^{-22}$ W recorded with the proof-of-principle receiver setup. The arrow points to the location of the insets. Top inset shows the signal with the background fit. Bottom inset shows the residuals from the background fit. The test signal is detected with a significance of $\gtrsim4.8\sigma$. Figure taken from \cite{Beaujean}.}
\label{fg:fakeaxion}
\end{figure}

Measurements have been performed with the proof-of-principle receiver chain in which a \lq\lq{fake axion}\rq\rq~signal at $18.4$~GHz with power of $\sim1.2\times10^{-22}$~W and bandwidth of $\sim10$~kHz is injected.~\footnote{The \lq\lq{fake axion}\rq\rq~signal is injected directly into the receiver without going through an antenna.} The total receiver noise temperature is $T_{rec}\approx5$~K as it results mainly from the input-referred noise temperature of the preamplifier. The background temperature is 4~K due to the liquid helium environment, making the system noise temperature $T_{sys}\approx9$~K. The fake axion signal could be detected~\cite{Beaujean}
%using the cross-correlation method 
with two days of measurement time with a significance of $\gtrsim4.8\sigma$. Fig.~\ref{fg:fakeaxion} shows the measured power spectrum which is dominated by thermal noise, and the normalized spectrum in which the excess from the \lq\lq{fake axion}\rq\rq~signal can clearly be seen. 

% system noise temperature
It has been demonstrated that the proof-of-principle receiver meets the requirements and yields $T_{rec}\approx5-6$ K. In order to be able to detect an axion signal of the benchmark power of $10^{-22}$ W within a few days measurement time, the noise temperature of the remaining system $T_{booster}$ should not exceed a few Kelvin. 
According to Kirchhoff's law of thermal radiation, the thermal noise of the booster is calculated by multiplying the physical temperature with its emissivity.
%, which is a characterization of the effectiveness of a given material in emitting thermal radiation and is the same as its thermal absorption coefficient. 
The emissivity of a perfect metal mirror would be zero and therefore zero noise temperature, while a lossy system such as the dielectric disc or a lossy metal mirror would have an emissivity between $0$ and $1$. By choosing a dielectric material with small $\tan\delta$ loss, the noise temperature of each dielectric disc can be made negligible. Therefore, in order to reduce $T_{booster}$ to $\sim2$ K, the entire booster should be enclosed in a helium cryostat, and low-loss metals and dielectric material should be used for the mechanical structure and discs, respectively.

\subsubsection{Quantum-limited linear amplifiers}
\label{sec:quantumamp}
It is evident from Eq.~\ref{eq:dickesmpl} that a low $T_{sys}$ is crucial for the success of \MADMAX. While HEMT preamplifiers could provide sufficiently low noise temperature, it is possible to lower the preamplifier noise temperature close to the theoretical lower limit, namely the so-called quantum limit:
\begin{equation}\label{eq:quantumlimit}
 T=h\nu/k_B,
 \end{equation} 
 where $h$ is the Planck constant and $\nu$ is the signal frequency. When $\nu=20$~GHz for example, the quantum-limited noise temperature is $T\approx1$~K -- significantly lower than the noise temperature of state-of-the-art HEMT amplifiers. In many cavity-based axion searches, near-quantum limit performance has been achieved by using superconducting Josephson junction or SQUID based parametric amplifiers. These amplifiers are built around the parametric interaction between three waves: signal, pump and idler. The nonlinear element which enables the parametric amplification is provided by a Josephson junction or SQUID, which has nonlinear inductance. However, such devices used by other axion searches are typically operated at $\lesssim10$~GHz with bandwidth less than a few MHz and low saturation power. In other words, quantum-limited amplifiers for \MADMAX~which work at $>10$~GHz with bandwidth $>50$ MHz and saturation power $\gtrsim-100$~dBm are yet to be developed.
 
Progress has been made to increase the bandwidth -- which naturally increases the saturation power -- by engineering a nonlinear transmission line using Josephson junction or SQUID metamaterial. Such devices are dubbed traveling wave parametric amplifiers (TWPA). For example, Ref.~\cite{bib:jtwpa} presents such a device made of an array of $>2000$ aluminum SQUIDs; it has a $3$~GHz bandwidth, $-102$~dBm 1-dB compression point and added noise near the quantum limit. Although the center frequency of the device is $\sim7$~GHz, experts agree that frequencies up to $40$~GHz -- roughly half of the superconducting gap of aluminum -- can be achieved by adjusting the SQUID array parameters. Plans have been put into place for the development of quantum limited parametric amplifiers for \MADMAX~led by the N\'{e}el Institute in Grenoble, France and the Walter-Mei{\ss}ner Institute in Garching, Germany. It is expected that the first working device in the range of $10$ to $15$~GHz will be delivered in 2020, while devices working for up to $40$~GHz could be developed in the next three years. In the meantime, MPP Munich is in the process of acquiring a dilution refrigerator (DR) in 2020, which is needed in order to provide the $\lesssim100$~mK temperature necessary for operating the amplifier. Such a setup would enable the characterization and testing of the device, as well as experimenting with methods with which it can be integrated into \MADMAX.

\subsubsection{Axion detection beyond 40~GHz}

\label{sec:highfreq}
The detection of weak signals beyond $40$~GHz is still a conundrum for the axion community. At this frequency range, even the quantum limit increasingly becomes a limiting factor for the sensitivity. To get around the quantum limit, many cavity experiments are actively pursuing single photon detection using technologies such as Rydberg atom microwave electrometry~\cite{bib:rydberg}. Aside from the intrinsic technological challenges, single photon detection would only be useful for \MADMAX\ if tunable filtering can be achieved between the antenna output and single photon counter input, given the higher ambient temperature and broader bandwidth compared to typical cavity experiments.

Linear amplification with near quantum limited noise could still be a viable option for \MADMAX~at higher frequencies. One possibility is using a SIS mixer to shift the axion signal to a lower frequency without introducing too much noise before amplifying it with a quantum limited amplifier. An SIS mixer which works below $100$~GHz with sufficiently low noise still needs to be developed. Alternatively, TWPA with Josephson junctions or SQUIDs made of materials with a higher superconducting gap such as Niobium could potentially increase the operating frequency to $\sim100$~GHz. Developing such devices will require more dedicated studies, different fabrication techniques and a longer time commitment. 

\section{Development of the \MADMAX{} booster}
The booster consists of a fixed copper mirror, the dielectric discs, the construction that keeps in place and moves the discs, and a system that allows to obtain the real-time position of the discs. It has to be operated at liquid helium temperature inside a high B-field, with a disc positioning accuracy of a few $\mu$m (see Sec.~\ref{sec:sim}). The maximum disc displacement is 1.2~m. There is no off-the-shelf solution for this long stroke positioning in a cryogenic environment. For several parts of the booster system, significant developments are required to meet these goals. 
%Also the overall design has to allow for the needed accurate disc positioning in the given surrounding.

\subsection{Booster design}  
\label{sec:booster}
A design study was performed by a company experienced in cryogenic engineering and drive concepts for cryogenic environment, JPE \cite{JPE:slit_unit1,JPE:slit_unit2}\footnote{https://www.janssenprecisionengineering.com/}. The study aimed at identifying the most promising mechanical concepts for precise movement of the booster discs. 
It is concluded from the study that 
%the design options presented in Fig.~\ref{fig:JPE-report} are feasible. 
a concept with moving motors on fixed guides  is considered to have lowest risk and cost. Concepts  where fixed motors push  the discs with moving rods which are mounted on sliding carriages were also investigated. The difference between the two variants is the temperature at which the motors are operated.    
%\begin{figure}
%    \centering
%    \includegraphics[width=0.95\linewidth]{FIGURES/JPE-report.png}
%    \caption{Concluding slide from the JPE design study report on the comparison of four drive concepts for cryogenic environment.}
%    \label{fig:JPE-report}
%\end{figure}
Another  concept has fixed motors moving a rod, but in this case the discs are directly fixed to the rods and moved by them without an intermediate carriage. This concept is deemed simple, but risky due to the unpredictable effect of misalignment. 

The booster design needs now to be further developed and visualized in 3D CAD on a system level in order to judge the interaction and space requirements of all relevant parts including, e.g., an actuator, guiding system, cooling solutions, an interferometer, etc. 
In the next step, the technical visualization of one selected concept will be commissioned to the JPE company. This will include the concept for a JPE motor, suitable for the application. The mechanical design will be, however, applicable to motors from other suppliers, as long as they meet the boundary conditions regarding dimensions, forces, tolerances, heat dissipation and stress resistance. These boundary conditions will be defined at the beginning of the project.

\subsection{Investigation of motors}  
\label{sec:motors}

Piezoelectric motors meet the majority of \MADMAX{} requirements. They can have high precision control, high force, low profile, low power consumption and quick response. They function in a vacuum under a strong magnetic field. Nonetheless, a few challenges have to be addressed for the adoption of piezo motors for \MADMAX.

%Piezo crystals in principle work at these cryogenic temperature. One of the main concern is thermal contraction of the materials. A few working examples exist already~\cite{}, yet not extensively tested at the scale of \MADMAX.

The performance of the piezoelectricity, especially for the large strokes needed, has to verified at the working temperature. The displacement of lead zirconate titanate (PZT) crystals -- the most commonly used piezoelectric crystal -- decreases rapidly at cryogenic temperature~\cite{piezo_temperature}. 
%At  \SI{4}{\kelvin}, the remaining displacement of the piezoelectricity is less than 10\,\% of the room temperature displacement. 
Consequently, the drive system has to be adapted for the low displacement. Alternative materials, such as electrostrictive single crystal, retain a better performance at cryogenic temperature and are being considered. Bipolar drive voltage can also increase the displacement, albeit at the cost of more complicated electronics.

The thermal load of the piezo motor has to be considered as well. 
%Higher power consumption during operation or due to the leakage through the drive cables require more cooling power. 
Each piezo motor dissipates about \SI{30}{mW} of power during its operation at 4\,K. This will be the dominant thermal load inside the cryostat during disc displacement. Dedicated thermal coupling structures, such as metal-wire brush or sliding contacts, can solve this issue. 

The high B-field does not interfere with the operational principle of the piezo elements. A few models, for example those used in NMR devices, are routinely operated in strong magnetic fields by simply replacing their magnetic materials with non-magnetic ones. However, few tests exist for operating piezo motors in a magnetic field as high as \SI{9}{T}. Dedicated tests will be carried out for \MADMAX{} in available test magnets (See Tab.\ref{tab:test_magnets}).
The reliability  of the  piezo  motors is an  important  criterion. The  final system includes 240 motors that have to operate reliably for years. A motor failure implies a technical stop of few weeks to open the experiment and extract the booster from the cryogenic vessel. Under this constrain, the motor failure rate must be limited to an acceptable level, and redundancy in the system has to be guaranteed.

The \MADMAX{} collaboration has identified several companies that deliver cryo-compatible piezo motors. One custom-made motor
has been tested inside a LHe cryostat in a \SI{5}{\kelvin} surrounding. The motor did work and a stroke of $\approx$\,\SI{10}{cm} was achieved without warming up the test surrounding significantly. However, the motor itself needs to be warmed up to $\approx$\,\SI{50}{\kelvin} before a significant movement could be reliably achieved. While this shows that piezo motors can in principle be used in the cryogenic surrounding, the thermal behavior of this motor type does not satisfy the requirements.

Alternatively, motors from two other companies are also being investigated. The company JPE has piezo motors in their product line that are rated for operation in 4\,K environment and below. %\footprint{https://www.janssenprecisionengineering.com/cryo-nano-positioning/}. 
An adaptation of one of these motors is being investigated in connection to the mechanical engineering study performed for the booster design as described in Sec.~\ref{sec:booster}.

\subsection{Disc material}	 
\label{sec:dielectrics}
Some single crystal materials are known to have the needed low dielectric loss of $\tan\delta < \num{e-4}$. 
The main candidate materials are \LALO{}~\cite{krupka1994,shimada2010,mazierska2004} and sapphire. \LALO{} is preferred due to its high dielectric constant of  $\epsilon\sim$\,\num{24}. Although measured values for the dielectric losses of both materials exist, they need to be evaluated at the frequencies relevant to \MADMAX{} and ultimately at cryogenic temperatures of about \SI{4}{\kelvin}. Additionally, at least for \LALO{}, it is known that the dielectric losses may vary largely between different samples/batches~\cite{krupka1994}, or even for individual samples, making it necessary to "screen" the stock material used to produce the discs in a kind of quality assessment.
Single crystals are not available with the required size, necessitating the tiling of the discs from smaller parts.
Alternatively, low dielectric-loss amorphous materials could be a solution, provided they can be produced in the needed size and with the needed accuracy.

\subsubsection{Alternative materials}
The tiling of the discs has an effect on the beam shape (see Sec.\ref{sec:sim}). Increasing the size of the elements from which the discs are tiled or even achieving disc production with the required size would reduce these effects. Also, mechanical accuracy and stability of the discs could be increased. 
%Current baseline design  material for \MADMAX{}, single crystal \LALO{} cannot be obtained as larger than 3 inch diameter disks. This size restriction stems from the high and uniform crucible temperature ($>\SI{2080}{\celsius}$)  required to grow a large diameter crystal. 
%In order to mitigate this, \LALO{} disks are stycast glued  to make a larger disc. 
Therefore alternative dielectric materials with adequate electromagnetic properties and bigger diameters are being researched. 
Also stratification of different materials is considered as option to increase the boost factor.
%behavior
%made into a large uniform mechanically stable disc but also possible combinations of materials that could yield the highest axion signal from the Booster. 

Low loss dielectric materials have been of great interest in industry and there are numerous of applications ranging from antennas for wireless communication to miniaturized electronics.  Ref.~\cite{dielectriclist} lists more than \num{4000} low-loss dielectrics among which candidate materials for the \MADMAX{} booster can be found. 
%We have recently found out that 
Furthermore, poly-crystalline \LALO{} can be formed into discs with  d$\approx\SI{50}{\centi\metre}$. It has been shown that these have very low losses at cryogenic temperatures~\cite{polylao}. However, the disc thickness might not be achievable that easily. Possibilities in this direction are presently being investigated.

%Therefore, to boost the chances of success of \MADMAX{}, an effort is being made to investigate the alternative materials where we identify, obtain and test the dielectric and mechanical properties of the potential candidate materials.

\subsection{Disc tiling}	 
\label{sec:tiling}
\begin{figure}[h]
\centering 
\includegraphics[width=0.5\linewidth]{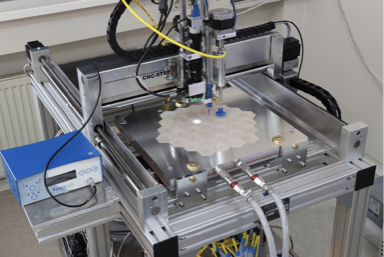}
\includegraphics[width=0.4\linewidth]{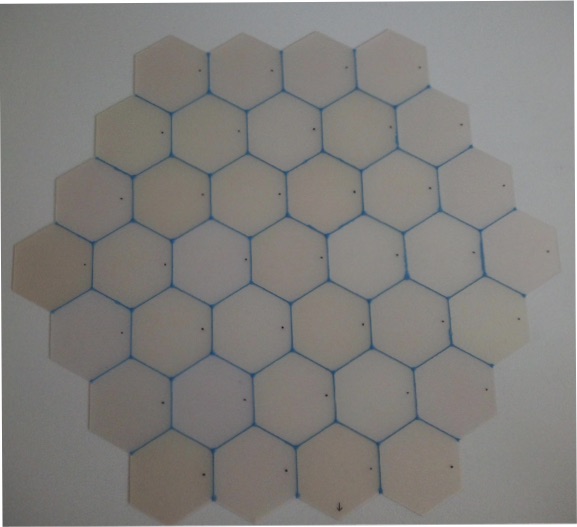}
\caption{Left) Disc tiling machine. Right) First glued LaAlO$_3$ disc for the prototype booster.}
\label{Fig:disc}
\end{figure}

With current technology \LALO{} can be grown  into single
crystals with sizes up to 3". Hence, the discs for the prototype booster with diameter 30~cm and later the final booster with diameter 1.2~m may have to be tiled from smaller pieces with the needed accuracy. 
A semi-automatic machine to position and glue hexagonal dielectric tiles has been assembled and commissioned. Stycast blue 2850ft is used for the gluing. Fig.~\ref{Fig:disc} (left) shows the machine being operated in the laboratory. The right photo shows the first successfully tiled LaAlO$_3$ disc using this machine. The technical specifications of the tiling machine are: plane flatness $<$~5~$\mu$m, gap between tiles 200~$\mu$m, homogeneous glue trace on all tail borders. After production the discs are characterized for flatness.   

\subsection{Disc characterization}
Simulation results suggest that the degradation of signal power from disc thickness variations can be tolerated if they are $\lesssim$10~$\mu$m or alternatively, if the disc thicknesses are  known to high enough accuracy (also after cooling down to LHe temperature) and the known variations can be taken into account when adjusting the disc positions.
%over length scales of a wavelength. 

\begin{figure}[h]
\centering 
\includegraphics[width=0.45\linewidth]{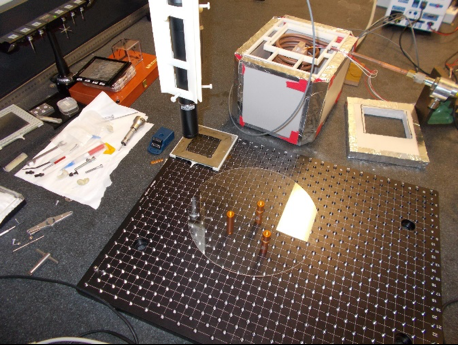}
\includegraphics[width=0.45\linewidth]{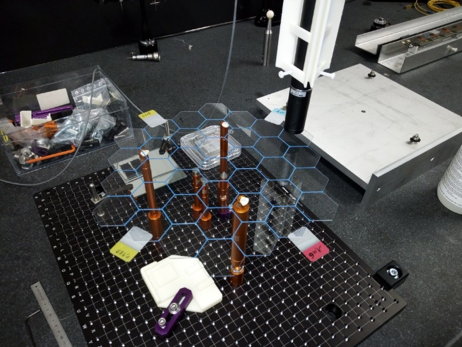}
\caption{3D optical setup used to measure a single sapphire disc of 20 cm diameter (left) and a tiled disc of 30 cm diameter made of glass (right).}
\label{Fig:planarity}
\end{figure}

Therefore it is interesting to precisely measure the surfaces and the disc thicknesses of all discs. 
%in order to control their planarity (flatness) and the disc thickness uniformity. 
This quality assurance of the disc manufacturing can be done with 3D-measurements. Two 3D tables are available to the \MADMAX{} collaboration, allowing to measure structures with size up to 1.6~m (also used for ATLAS pixel modules measurements). The machines are equipped with contact and non-contact sensors (optical pen and laser scanner). Measurements of disc surfaces can be performed both in vertical and horizontal positions, allowing them to check the impact of gravity on the disc shapes.  Measurements have already been performed on discs used for the first proof of principle measurements and on glass-discs tiled with the setup described in Sec.\ref{sec:tiling}. 
The setup is shown in Fig.~\ref{Fig:planarity}. 

A few thousand points were taken per disc side within 1 hour. The precision of the measurement setup has been demonstrated with an uncertainty of a few $\mu$m. The results of the first scans show that the tiled disks do not yet meet the specifications.
At this stage, the measurements allow to control the quality of the produced discs and helps to improve the production process.
These measurements will also be used for simulating the impact of surface unaccuracies on the boost factor.
%in the \MADMAX{}beam propagation simulation tool in order to check their impact on the boost factor.

%%%%%%%%%%%%%%%%%%%%%%%%%%%%%%%%%%%%%%

\subsection{Disc positions control system - interferometer}  
\label{sec:interferometer}
%The boost factor, and hence the sensitivity to an axion signal, is defined by the spacing between the discs. 
%Simulations have shown that the boost factor functions start to deviate significantly when the spacing vary by more than a few µm. 
The group delay of a microwave signal injected into the booster reveals information about the disc spacing.  The result of the measurement does, however, neither give the exact position of all discs nor the information on the tilt or unevenness of the disc surfaces.
It is therefore helpful to measure the spacing (or disc positions) and tilt independently with an accuracy of at least  $ 1\mu m$.

%The accuracy of this method is, however, only on the order of a few $10 \mu m $ and decreases rapidly with an increasing number of discs. 

%A market search has shown that due to t and 
For the required accuracy in the harsh environment, only optical measurements seem feasible. 
Two options have been studied in detail out of which one is considered for integration into the prototype: A precision laser interferometer from the company HighFinesse\footnote{https://www.highfinesse.com/en/}. The proposed interferometer provides an absolute measurement of the disc position with accuracy of $< \mu m$, allowing to determine the tilt of the discs with sufficient accuracy if three points per discs are measured. The system can provide real-time feedback at a rate of 10\,Hz.
The draw-back of this solution is that it requires special mirrors on 
each disc and an optical system inside the cryostat vessel.
%and transparent windows to feed the laser beams into the cryogenic vessel.
%low-coherence infra-red interferometry by the company SMARTEC and a newly developed solution based on precision laser interferometry 
%by the company HighFinesse, a spin-off from the quantum optics group at the University T\"ubingen.

%Currently the favoured solution is the one from HighFinesse as it provides the better accuracy ($< \mu µm$), works indpendently of the absolute distance between the discs, and can provide real-time feedback. 

So far a one-channel system was successfully tested at room temperature and in air to demonstrate the accuracy and stability. 
%In the prototype setup it is planned to measure each disc at 3 positions so a possible tilt of the disc can be measured. 
A 6-channel setup to determine the position of two movable discs is presently being commissioned. 
It will be also used to evaluate the robustness of the system against external influences and to gain operation experience.% (e.g. how frequently the system needs to be realigned, etc.).

A fallback option to the HighFinesse system is a commercially available interferometer offered by the company SMARACT \footnote{https://www.smaract.com/careers}, which provides similar precision but only relative position measurement.

An alternative option is a low-coherence infra red interferometer that relies on the reflectivity of the discs themselves, hence avoiding the need for mounting of retro-reflective mirrors. 
Such an interferometer has been integrated into the proof of principle booster setup (see Sec.\,\ref{sec:pop:tuning}). 
Initial tests showed an accuracy of a few $\mu m$. 
It was also  demonstrated that up to five sapphire discs could be measured with a single IR beam. 
This system cannot be used for \MADMAX{} in its current configuration due to the long measurement time of up to 10~s and due to the limited measurement range of $\lesssim$\,10\,cm. 
Extending this range and speeding up the measurement would require significant R\&D.

\section{The \MADMAX{} prototype}
\label{sec:prototype}
The \MADMAX{} prototype is a down-scaled version of the final experiment with the specific task of testing all technological components of \MADMAX{}, establishing the data taking procedure and potentially acquiring the first physics results on ALPs and hidden photon searches with this technology.  \\
The down-scaling is by a factor of 1/10 in area of the discs, 1/4 in the number of discs and 1/6 in the magnetic field strength. The \MADMAX{} prototype has the same structure and components as presented in section~\ref{sec:concept} but reduced overall size. A cryogenic vessel containing the prototype booster and the optical system is being designed to fit the open bore of the MORPURGO magnet at CERN. 
The prototype design is flexible enough to account for changes in the booster design to accommodate for the shortcomings identified during commissioning. The feed-through to the receiver is also kept flexible to offer the possibility of testing various receiver concepts.
Details of these components are presented in the following sessions. 

\subsection{Prototype booster design}
\label{sec:booster_design}
\begin{figure}
    \centering
    \includegraphics[width=0.45\linewidth]{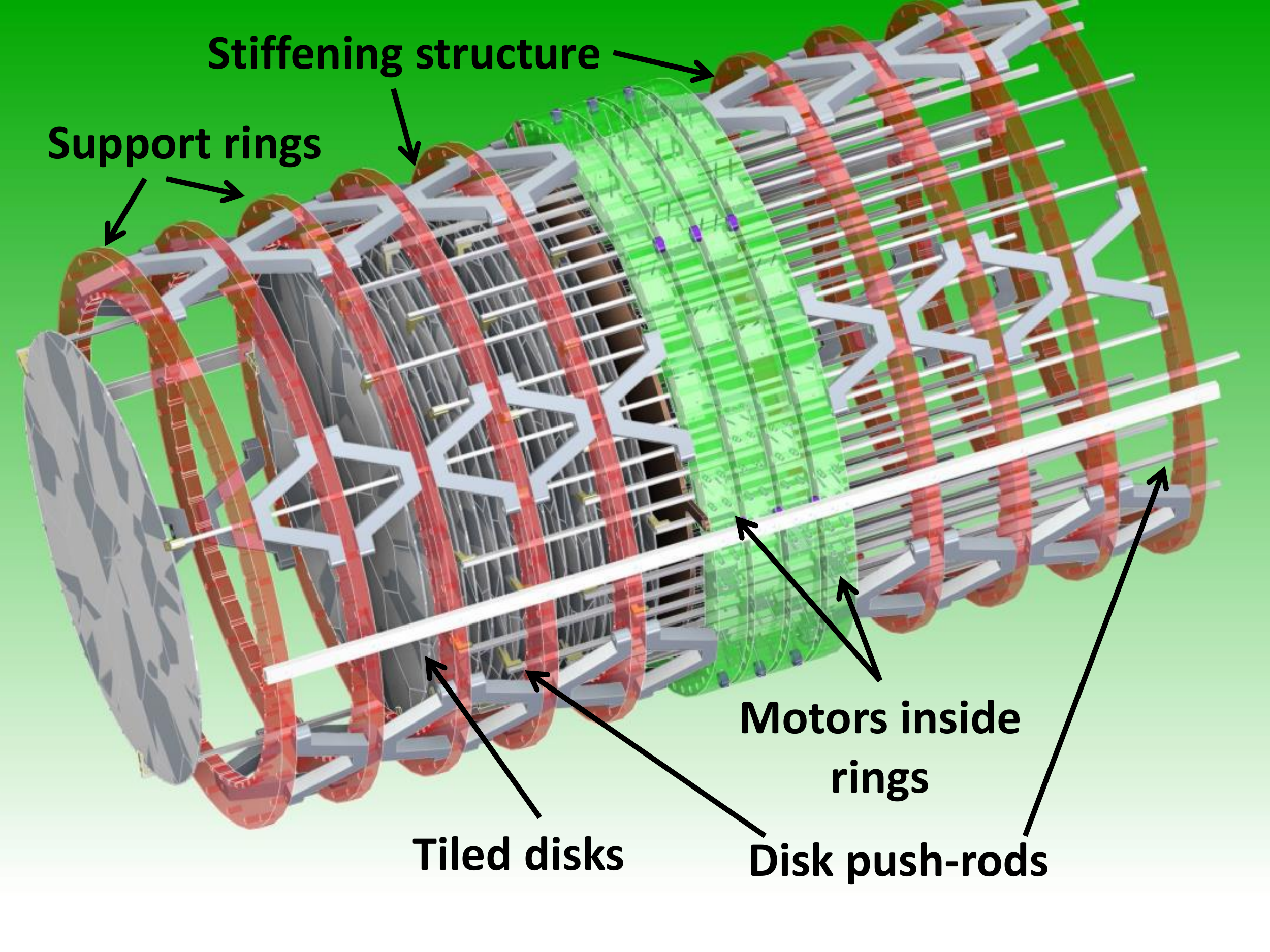}
    \includegraphics[width=0.45\linewidth]{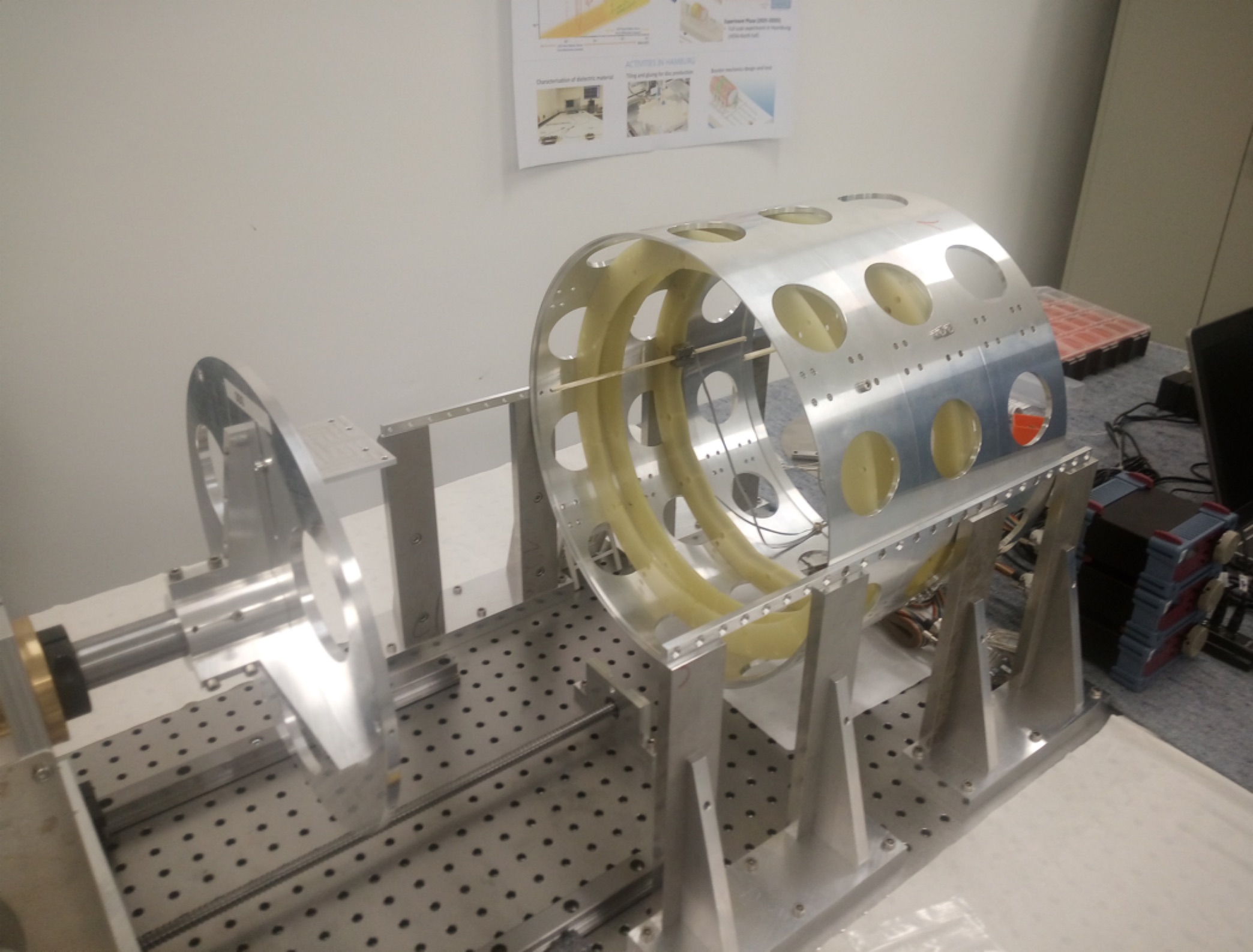}
    \caption{Left) Technical design of the mechanical structure for the prototype booster. Right) Pre-series production of the mechanical parts assembled for tests in the SHELL laboratory in Hamburg. }
    \label{fig:prototype_booster}
\end{figure}

The design of the booster prototype is being investigated. The structure shall support 20 discs of 30~cm diameter, the motors necessary to move them and the system to control the motors. It shall provide the sufficient rigidity to obtain the required alignment tolerance of the discs.
As discussed in section~\ref{sec:booster}, a few design options have been evaluated by the collaboration together with the company JPE. Two of these are being followed up. 
The first design is based on the idea that stationary piezo motors are driving the rods connected to the individual discs. Each disc is positioned by moving three rods, each driven by an individual piezo motor. A CAD drawing of this option is shown in Fig.~\ref{fig:prototype_booster} (left).  

This structure is being commissioned in the SHELL laboratory in Hamburg. A pre-series needed to develop the final concept is depicted in Fig.~\ref{fig:prototype_booster} (right). The mechanical parts for this pre-series booster are being machined and  assembled. A commissioning period will follow, where the required positioning precision of better then ~10~$\mu$m will be demonstrated.

The second design, believed to be the most promising by the JPE study, has motors integrated directly onto the disc holder system. The full implementation of this design is being worked on and will be available in early 2020. 

Final production drawings for the prototype booster will be available mid-2020. 

\subsection{The cryostat vessel}
For the \MADMAX{} prototype a cryostat is needed to host the prototype booster as well as the focusing mirror, and to provide an interface between the antenna and receiver system. 
It will have an inner diameter of \SI{750}{\milli\metre} allowing for installation of the focusing mirror. The outer diameter is limited by the constraints of the MORPURGO magnet at CERN (see section~\ref{sec:morpurgo})  and the foreseen place for commissioning of the whole prototype system, the SHELL laboratory. The cryostat vessel also includes the various interfaces to the subsystems of the prototype, e.g. the prototype booster including possibly a laser interferometer and the focusing mirror. These interfaces 
%are not only the mechanical mounts but also 
include electrical (and optical) feed-throughs as well as thermal couplings. To allow for flexibility and versatility and simpler commissioning, initially the antenna and receiver system will be placed at room temperature outside of the cryostat. The signal will be coupled to the receiver through a window. Later this will be replaced by a dedicated feed-through.

\begin{figure}
    \centering
    \includegraphics[width=.45\textwidth]{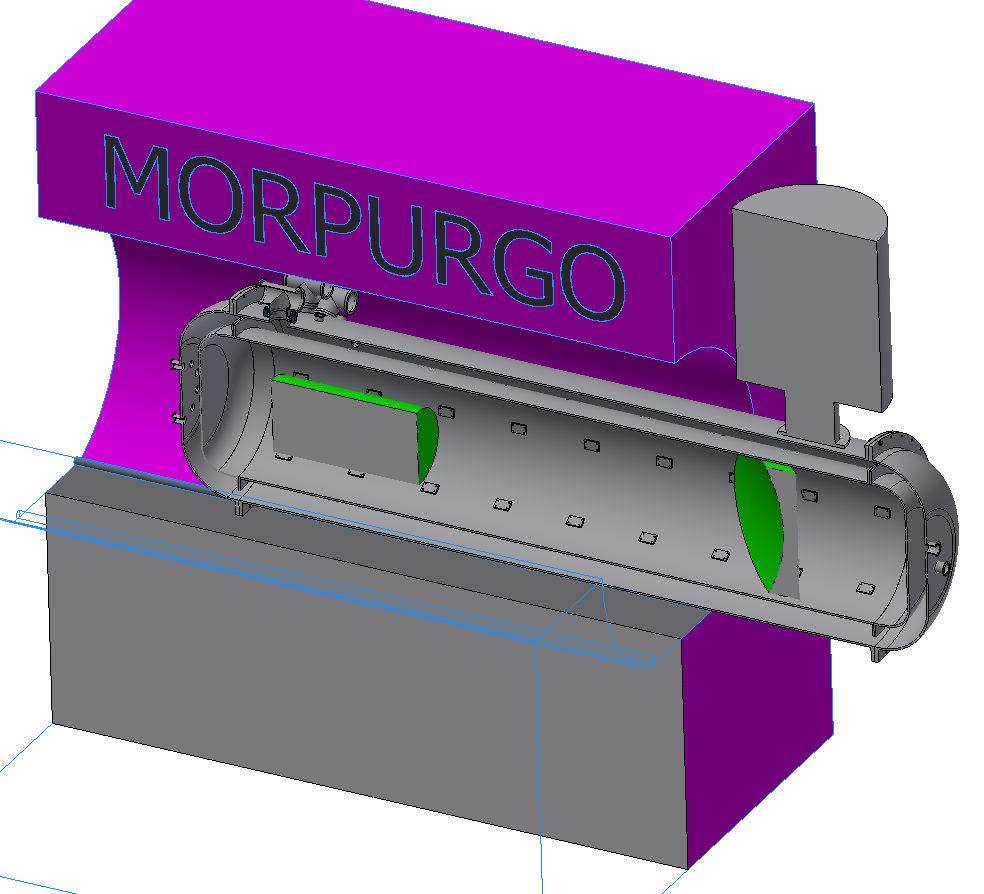}
    \includegraphics[width=.45\textwidth]{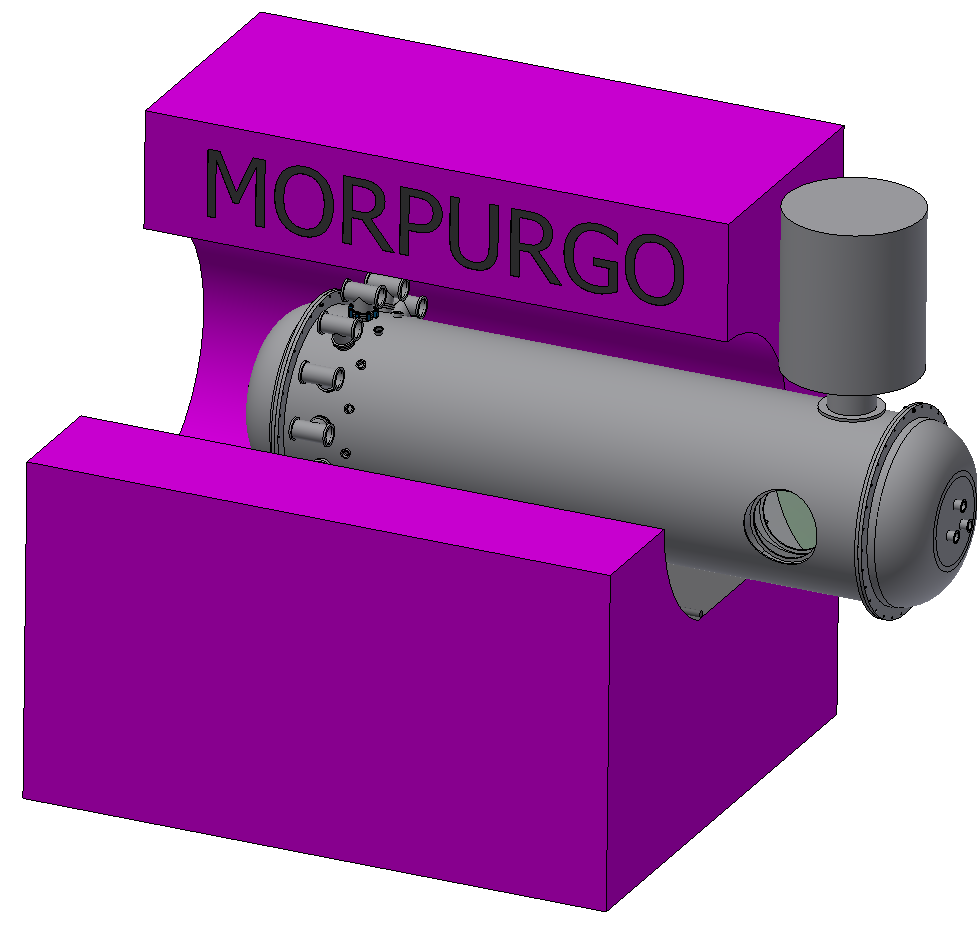}
    \caption{Sketch of the \MADMAX{} prototype cryostat vessel.}
    \label{fig:mpbv-sketch}
\end{figure}

A sketch of the cryostat vessel (placed inside the MORPURGO magnet, see section~\ref{sec:morpurgo}) can be found in Fig.~\ref{fig:mpbv-sketch} which shows the rough layout.
%and the dimensions already defined.

A first design study for the cryostat vessel has been started with Bilfinger Noell GmbH, explicitly looking into different cooling concepts: e.g. cryocoolers vs. LHe from dewars. The outcome of this design study should lead to a decision on the cooling concept in December 2019  so that a detailed design could be ordered in the context of a staged contract with Bilfinger Noell GmbH. This will be followed by the production of the cryostat vessel, to be delivered in March 2021 to the SHELL laboratory.

\subsection{Measurements in B-field}
\label{sec:morpurgo}
It is essential to study the behavior of the individual mechanical components inside a strong magnetic field.
Measurements are foreseen in different magnets available at DESY Hamburg, University of Hamburg and at CERN. The important parameters of the available magnets are listed in Tab. \ref{tab:test_magnets}. Two magnets with small usable aperture will be available in Hamburg for testing of individual components like the motors, dielectric materials, etc.

For testing the full prototype the MORPURGO dipole magnet \cite{madmax_morpurgo} at CERN could be available during the SPS shutdowns each year from December to April starting from December 2021. It has an aperture of 1.6\,m and a usable cross section of roughly 1.45\,m. After clarification of the feasibility with CERN representatives, a letter of intent to the CERN SPSC for the usage of the magnet has been handed in \cite{MORPURGO-LoI}. It is expected that the project can be presented to the CERN SPSC in early 2020.
For the case of a positive decision the \MADMAX{} collaboration can expect the support from the CERN engineering, electrical and technical departments during the installation of the setup in the north area.

\begin{table}
    \centering
    \begin{tabular}{l|l|l|l}
      Magnet   & ALPs test      & MORPURGO        & WISP-MAG\\
      \hline
      Location & DESY hall west & CERN north area & Uni. Hamburg\\
      Magnet type & Dipole      & Dipole          & Solenoid\\
      B$_{max}$ [T]    & 5.7       & 1.6  (1.9)       &  14  \\
      Usable aperture [m] & 0.055 & 1.45          & 0.150 \\
    Usable length [m]     & 8.8   & $\sim$\,1  & 0.01  \\
    Availability          & After 2020            & CERN SPS shutdowns &  from summer 2020 \\
    Access                & 4\,K cold bore & warm bore & warm bore \\
    Sample preparation time & $\sim$\,3 days & - & - \\
    Max dB/dt [T/s]         & 25                & $<$\,0.1  & t.b.c.\\
    \end{tabular}
    \caption{List of available magnets for testing prototype and individual components}
    \label{tab:test_magnets}
\end{table}

\subsection{Physics measurements}
Provided that the tests in the magnetic field of MORPURGO are successful and a reliable understanding of the booster optical behavior can be achieved, it is planned to use the prototype inside MORPURGO to produce first competitive limits for ALPs in the  mass range around $\approx$\,100\,$\mu$eV.
A sensitivity to the ALPs photon coupling (5\,$\sigma$ signal above system noise) down to $\approx$\,10$^{-12}$~GeV$^{-1}$ over a mass range of $\approx$\,5~$\mu$eV can be achieved using the following assumptions (see Fig.~\ref{fig:sensitivity_plot}, cyan):
 a sensitivity of the receiver as described in section~\ref{sec:hemt}, i.e. a system noise temperature of T$_{sys}\,\approx$\,8\,K; a coupling efficiency between the axion field and the receiver 50\,\% as compared
to the 1D calculations (loss effects, antenna coupling);
 %a coupling efficiency of the receiver to the antenna and thus to the booster signal of 50\,\%;
 a readjustment time of the booster discs between different optimized scanning ranges of 1 hour; and a scanning time of 90 days; smallest achievable boost-factor bandwidth of 50 and 20\,MHz (upper and lower line, respectively).
 Even without a magnet previously unexplored range in mass for hidden photons can be probed. 
 
 % Time and resources plan has been deleted for public release

\section{Summary and conclusions}

The \MADMAX{} collaboration is devoted to the challenging task of detecting the dark matter axion in the mass range around 100\,$\mu$eV.
The experiment will be based on the concept of the dielectric haloscope.

The basic principle of the boost of axion to photon conversion has been scrutinized in detail, including simulations of  3D effects.
These investigations show that \MADMAX{} could indeed reach the sensitivity needed to discover axion dark matter. While challenging, the production of the setup with the needed accuracy seems feasible.

The success of the project depends on the availability of a magnet with high enough B-field over a large enough  volume. There has been significant progress with respect to the development of such a magnet. The two innovation partners of the \MADMAX{} collaboration have independently concluded that it is feasible to build a magnet with the given specifications. They now cooperate on the project and are developing a common magnet design. The \MADMAX{} collaboration expects to have available a magnet that meets the demanding requirements by 2025.

The other main challenge of calibrating the boost factor and minimizing the uncertainties is being tackled by investigating several methods for verifying the understanding of the RF response of the system. 

The development of the different components of the setup are ongoing. As a first step towards realization, a down-scaled prototype experiment will be built. It is expected that this setup will be available in 2021. It will be used for extensive testing and understanding of the mechanical principles, as well as the RF behavior of such a system. Measurements with this prototype in an available dipole magnet - MORPURGO at CERN - will also be performed.
Such a prototype could be sensitive to an unexplored parameter range for ALPs and hidden photons and allow for first particle physics publication.

Preparations of the required cryogenic infrastructure to host \MADMAX{} at DESY Hamburg, like the needed cryogenic lines, are already ongoing.  

The \MADMAX{} collaboration is confident that it has identified the main challenges of the project, and that it has developed suitable strategies to solve them.

%We ask the DESY PRC to endorse the scientific relevance of the experiment and the plans of the collaboration for the next phase.

\section*{Acknowledgements}
This work is supported by the Deutsche Forschungsgemeinschaft under Germany’s Excellence Strategy - EXC 2121 Quantum Universe - 390833306 and under project number 441532750.

%% file: ExecutiveSummary.tex
\newpage
\section{Executive summary}
%The axion is an elementary particle candidate to solve the strong CP problem. It turns out to also be a very well motivated cold dark matter candidate. 
The axion, originally introduced to solve the strong CP problem, is an excellent dark matter candidate. 
%The dielectric haloscope has been proposed as a detection method for dark matter axions in the theoretically well motivated mass range of $40-400\,\mu$eV.
The dielectric haloscope is a promising technique for dark matter axion detection in the theoretically well motivated mass range of $40-400\,\mu$eV.
%compatible with the post inflationary Peccei-Quinn symmetry breaking. 
The detection principle is based on the conversion of axions into photons in a strong magnetic field. The interface between vacuum and dielectric materials is used as a conversion surface. The tiny power produced in the conversion is ``boosted'' by coherent photon emission from multiple dielectric disk surfaces combined with constructive interference effects when placed in front of a mirror. The ``booster'' concept has already been scrutinized in some detail~\cite{madmax_prl,madmax_foundations,Ioannisian:2017srr,Millar:2017eoc}.

The \MADMAX{} project is divided into phases:
\begin{itemize}
    \item \textbf{Feasibility} of detector components; of magnet procurement; simulation studies including determination of required mechanical precision.
    \item \textbf{Prototype} of down-scaled booster system and test in existing magnet. 
    \item \textbf{Experiment} construction and commissioning of final booster system, receiver and magnet.
    \item \textbf{Axion search} scanning of the relevant parameter space.
\end{itemize}

The {\bf feasibility phase is concluded}. A feasibility study on the dipole magnet  has been completed with two magnet suppliers within an innovation partnership - a novel tendering tool of the European Union \cite{madmax_innovation_partnership}. It was shown by the two partners independently that it is possible to build a magnet with the needed specifications 
%in block design based on NbTi superconductor and a peak field of 9~T  
within realistic time and budget constraints. An external expert committee has evaluated these studies and confirmed the outcome, and has encouraged the innovation partners to proceed towards a realization of the magnet.
The two innovation partners are now working together in a novel form of academic-industrial partnership. 
%Presently a conceptual magnet design is being detailed and a realistic price tag and time schedule for the overall magnet is being worked out.
%Some details on the magnet project, the concepts and design as well as plans are given in Sec. \ref{sec:magnet}. 

Simulation studies of the experiment have been carried out using tools modified to take into account an axion source term. 
These suggest that an amplification factor of $\gtrsim 10^4$ for axion to photon conversion  can be obtained by the booster system.
%, and is reported about in Sec.~\ref{sec:sim}. 
The feasibility of the booster concept is being experimentally tested using a proof of principle setup. %the results are discussed in Sec.~\ref{sec:meas-setup}. 

%Progress has been made on understanding the mechanics and the electromagnetic system have . 
%No show stoppers on the mechanical and electromagnetic system have been identified.

The collaboration is currently {\bf entering the prototype phase}, which will demonstrate the experiment technology in all its aspects, and will provide a design scalable to the final experiment. It is planned to test the prototype detector inside the MORPURGO dipole magnet located at CERN. First physics results in terms of hidden photon and Axion Like Particle (ALP) searches will be achieved.

All aspects of the prototype are presently being investigated to identify the optimal technological solutions. No show stoppers have been identified. It is expected that all required inputs for the conceptual design report for the prototype will be ready in 2020. 

In parallel, the final experimental site, the HERA North hall at DESY is being refurbished to host the \MADMAX{} experiment. A radio-quiet laboratory (SHELL) dedicated to commissioning of the prototype has recently been brought into operation at the University of Hamburg.

%The in-depth review of the project by the panel of experts of the DESY PRC is timely. The collaboration is confident about the feasibility of the project and about the strategy to tackle the technological challenges. It further believes that the most critical aspects of the project have been identified and will be properly addressed in the prototype phase. 

%The \MADMAX{} collaboration will strongly benefit from the critical review of the committee and asks for the endorsement of
%\begin{itemize}
%    \item the scientific relevance of the experiment,
%    \item the plans for the prototype phase.
%\end{itemize}

%This report first summarizes the theoretical motivation of the \MADMAX{} project. The basic concept of the experiment is introduced and a coherent strategy to tackle the individual technological challenges is presented.
%The technological feasibility of the magnet is then reviewed. 
%Details on the individual technological challenges are given. 
%The next plans with the prototype are discussed and the time and budget planning is described.